\pdfoutput=1
\documentclass[pdflatex,sn-nature]{sn-jnl}

\usepackage[american]{babel}
\usepackage[T1]{fontenc}
\usepackage[utf8]{inputenc}
\usepackage{amssymb}

\usepackage{graphicx}
\usepackage{amsmath}
\usepackage{siunitx}
\usepackage{xcolor}
\usepackage{orcidlink}
\renewcommand{\orcid}[1]{\,\orcidlink{#1}}
\DeclareSIUnit{\electron}{e^-}
\DeclareSIUnit{\ADU}{ADU}

\DeclareUnicodeCharacter{2212}{-}

\begin{document}

\title[Supercritical light collection in NA\,$>$\,1 fiber bundles]{Supercritical light collection in ultra-high-numerical-aperture fiber bundles for wide-field quantum magnetic endoscopy}

\author*[1]{\fnm{Mariusz} \sur{Mrózek}\orcid{0000-0003-1824-8357}}\email{mariusz.mrozek@uj.edu.pl}
\author[1,2]{\fnm{Saravanan} \sur{Sengottuvel}\orcid{0000-0002-8984-735X}}
\author[3]{\fnm{Adam} \sur{Filipkowski}\orcid{0000-0002-1027-6751}}
\author[3]{\fnm{Grzegorz} \sur{Stępniewski}\orcid{0000-0003-2286-6764}}
\author[3]{\fnm{Dariusz} \sur{Pysz}}
\author[3]{\fnm{Ryszard} \sur{Buczyński}\orcid{0000-0003-2863-725X}}
\author*[1]{\fnm{Adam M.} \sur{Wojciechowski}\orcid{0000-0003-1805-6718}}\email{a.wojciechowski@uj.edu.pl}

\affil[1]{\orgname{Institute of Physics, Jagiellonian University}, \orgaddress{\street{Łojasiewicza 11}, \postcode{30-348}, \city{Kraków}, \country{Poland}}}
\affil[2]{\orgname{Jagiellonian University, Doctoral School of Exact and Natural Sciences}, \orgaddress{\street{Łojasiewicza 11}, \postcode{30-348}, \city{Kraków}, \country{Poland}}}
\affil[3]{\orgname{Centre of Excellence for Structured Fiber Optics and Applications, Faculty of Physics, University of Warsaw}, \orgaddress{\street{Pasteura 5}, \postcode{02-093}, \city{Warsaw}, \country{Poland}}}

\abstract{\unboldmath Nanoscale emitters near a high-index surface send most of their light into supercritical angles, which conventional fibers cannot accept and only immersion optics normally collect. Here we show analytically that fibers with a numerical aperture exceeding the refractive index of the emitter's environment, $\mathrm{NA} > n_\mathrm{env}$, capture this near-field emission directly as light guided along the fiber, without nanofabricated resonators or single-emitter positioning. Specifically, for nanoscale emitters such as nitrogen--vacancy (NV) centers in nanodiamonds on an NA\,=\,1.15 fiber in air, the model and full-wave simulations give a coupling efficiency of 0.48, nine times that of a conventional high-NA fiber (NA\,=\,0.5), with about 40\% of the guided light supercritical. Experiments confirm the analysis: back-focal-plane imaging and magnetic-resonance measurements show this light at the remote fiber end, where immersion readout increases the signal 1.9-fold. Using an imaging fiber bundle of 830 such cores, each acting as one pixel, we demonstrate microwave-free, wide-field magnetic endoscopy. The rule $\mathrm{NA} > n_\mathrm{env}$ applies to near-field collection from any nanoscale emitter.}

\maketitle

\section*{Introduction}

Efficient coupling of light from nanoscale emitters into well-defined optical channels is a central challenge in nanophotonics and quantum sensing \cite{Aharonovich2016, Lodahl2015}. Nitrogen--vacancy (NV) centers in diamond \cite{Doherty2013, Rondin2014, Degen2017} are a prominent example: they combine optical spin readout with long spin coherence at room temperature \cite{Ju2021}, and dense NV ensembles in bulk crystals and nanodiamonds (NDs) are widely used for magnetometry \cite{Acosta2009}. Their readout is limited by photon shot noise, so every photon that is not collected costs sensitivity. However, a large fraction of the emission lies outside the acceptance cone of conventional collection optics \cite{Lukosz1977_I, Lukosz1977_II, Lukosz1979_III, Christinck2020}.

This limitation originates from the electrodynamics of a dipole near a dielectric interface \cite{Lukosz1977_I, Lukosz1977_II, Lukosz1979_III, Novotny2012}. An emitter is surrounded by a near field that does not propagate in air but decays within a fraction of a wavelength. When a dense medium enters this near field, part of it turns into light propagating inside the medium---the same tunneling that frustrates total internal reflection when two prisms are brought within a wavelength of each other. This light carries transverse wavevectors $k_\parallel$ larger than the free-space wavevector $k_0$. Throughout this work, we use the normalized transverse momentum $q = k_\parallel/k_0$, which equals $n\sin\theta$ for a plane wave traveling at an angle $\theta$ in a medium of index $n$ and is conserved across a flat interface, and we call components with $q > 1$ supercritical. They are evanescent in air, but propagate in a medium with $n > q$, where they appear beyond the critical angle as supercritical-angle fluorescence (SAF). SAF is routinely collected with immersion objectives and is exploited in microscopy and emitter localization \cite{Oheim2020, Winterflood2010}, but this requires bulky optics in immersion contact with the substrate.

Two families of structures address the extraction problem. Monolithic elements fabricated around an emitter---solid immersion lenses, circular bullseye gratings, and parabolic reflectors---redirect the emission into low-NA optics with collection efficiencies of 70--90\% \cite{Hadden2010, Li2015, Wan2018}. Waveguide interfaces---semiconductor nanowires, photonic crystal waveguides and tapered nanofibers carrying single nanodiamonds---channel the emission into a single guided mode, with coupling efficiencies above 0.9 in the best devices \cite{Claudon2010, Arcari2014, Lodahl2015, Schroder2012, Sheremet2023}. Diamond nanopillars, which serve as the tips of atomic force microscopy (AFM) probes in scanning NV magnetometry, belong to the second family: the high diamond--air index contrast guides the emission of an NV center inside the pillar towards the collection optics \cite{Babinec2010, Maletinsky2012, Momenzadeh2015, Appel2016, Hedrich2020}. Both families rely on nanofabrication or positioning of individual emitters and address one emitter at a time, which excludes parallel, wide-field readout. Conventional optical fibers are robust and scalable, but their numerical aperture ($\mathrm{NA} \le 0.5$) restricts collection to a few percent of the dipole emission. They have been used for remote single-point NV magnetometry \cite{Fedotov2014, Maayani2019, Czarnecka2022} and, with a fiber-integrated diamond probe, for scanning magnetic-field imaging \cite{Fedotov2016}. Imaging fiber bundles with nanodiamond-coated facets and $\mathrm{NA} \approx 0.5$ have been used to record optically detected magnetic resonance (ODMR) signals through individual fibers and to detect magnetic-field differences and gradients across the bundle facet \cite{Wojciechowski2019, Jani2023}; these demonstrations relied on microwave-driven ODMR, and the role of the fiber NA in collecting near-field emission was not analyzed.

In a waveguide, the numerical aperture is a material parameter set by the core--cladding index contrast \cite{SnyderLove1983, Okamoto2006}: 
\begin{equation}
\mathrm{NA} = \sqrt{n_{\text{core}}^2 - n_{\text{clad}}^2},
\label{eq:NA}
\end{equation}
and can exceed unity even when the facet is in air; soft-glass fibers and imaging bundles of this kind have recently been developed \cite{Bijoch2023}. Such a fiber never shows its full NA as a cone of light in air: guided light with $q > 1$ is totally reflected at a dry output facet, much as light from outside Snell's window stays trapped below the surface of water. At the input facet, however, the same guided modes with $q > 1$ can be fed by the near field of an emitter, whose evanescent components become propagating waves in the core. For near-field light, the fiber is thus a one-way door: it accepts more than it can release into air. The relevant comparison is therefore not with unity but with the refractive index $n_\mathrm{env}$ of the emitter's environment: only for $\mathrm{NA} > n_\mathrm{env}$ does the acceptance of the fiber reach beyond the light line of the environment and include supercritical emission. Such a fiber transfers the principle of the high-contrast nanopillar to a macroscopic, multimode waveguide, with the emitter in the near field of the facet rather than inside the guide.

Here we show that step-index fibers with $\mathrm{NA} > 1$ occupy the intermediate regime between unguided SAF and single-mode nanostructures. Using NV centers in nanodiamonds as a model system, we (i) quantify this capture with an analytical momentum-space model and full-wave simulations, which give a coupling efficiency of 0.48, nine times that of a conventional high-NA fiber (NA\,=\,0.5); (ii) identify $\mathrm{NA} > n_\mathrm{env}$ as the design rule; (iii) show with back-focal-plane imaging and with dry and immersion readout that the captured light reaches the far end of the fiber; and (iv) use an 830-core NA\,=\,1.15 imaging bundle as a lensless endoscope for wide-field magnetic imaging, in which every core acts as an independent near-field collector. The endoscope is read out all-optically with a near-zero-field cross-relaxation protocol \cite{Dhungel2024, Sengottuvel2025, Dhungel2026}, which requires no microwave instrumentation near the sample and works with randomly oriented NDs.

\section*{Results}

\subsection*{Momentum-space picture of near-field capture}

\begin{figure*}[!tbp]
    \centering
    \makebox[\textwidth][c]{\includegraphics[width=145mm]{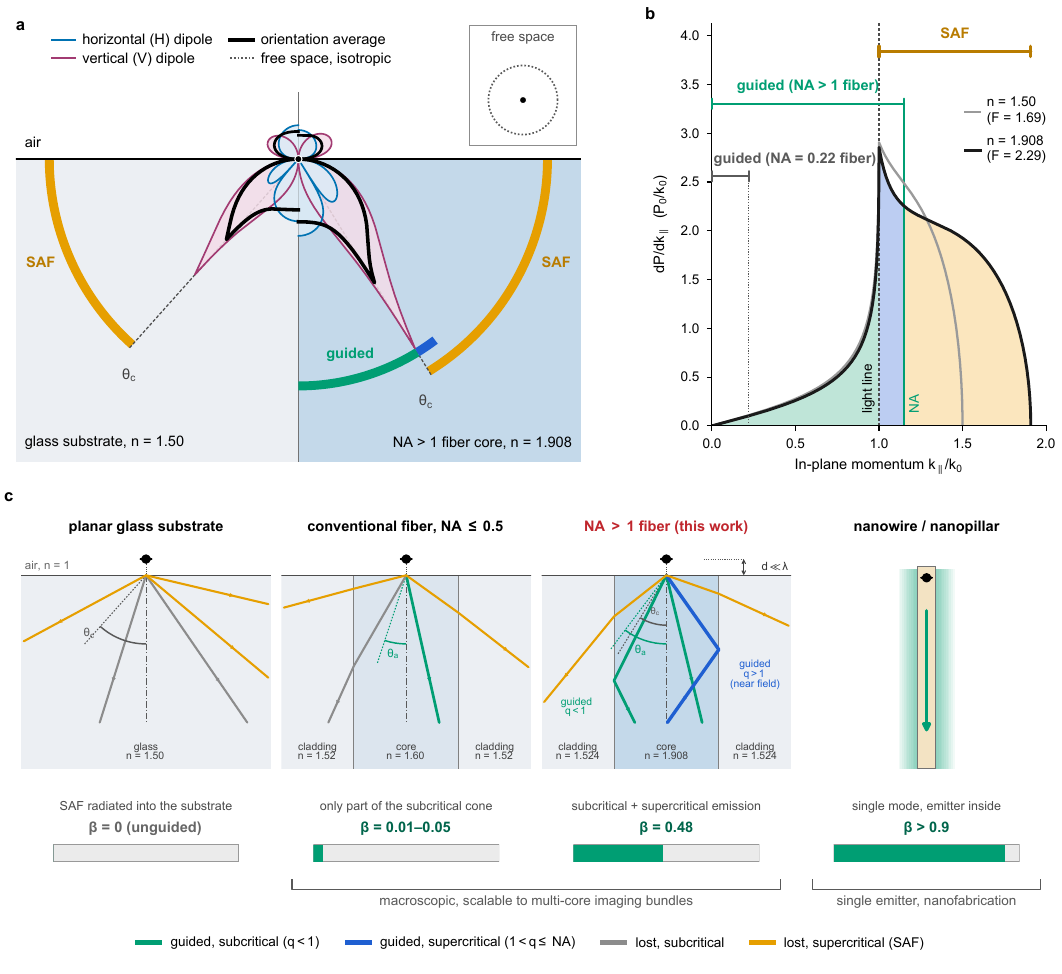}}
    \caption{\textbf{Supercritical-angle emission and its capture by an NA\,$>$\,1 fiber.}
    \textbf{a},~Angular emission of a dipole at a distance $d \to 0$ from the interface between air and ordinary glass (left, $n = 1.50$) or the core of an NA\,$>$\,1 fiber (right, $n = 1.908$). Thin lines: horizontal (H; cross-section in the plane of the dipole) and vertical (V) dipoles; thick line: orientation average. The radius is proportional to $(\mathrm{d}P/\mathrm{d}\Omega)^{1/2}$, where $P$ is the emitted power and $\Omega$ the solid angle, on a common scale with an isotropic emitter in free space (inset, dotted circle). Outer arcs, colored as in \textbf{c}: supercritical-angle fluorescence (SAF) range beyond the critical angle, $\theta > \theta_\mathrm{c}$, and acceptance cone of an NA\,=\,1.15 fiber.
    \textbf{b},~Emission into the dense medium ($n = 1.908$, black; grey: ordinary glass, $n = 1.50$) per unit normalized transverse momentum $q = k_\parallel/k_0$, relative to the free-space power $P_0$, for an orientation-averaged dipole at the interface. Shaded: emission guided by the NA\,=\,1.15 fiber below ($q < 1$) and above ($1 < q \le \mathrm{NA}$) the light line, and emission beyond the fiber cutoff. Bar: acceptance of a conventional NA\,=\,0.22 fiber. $F$, total emitted power relative to $P_0$ (Purcell factor).
    \textbf{c},~Coupling regimes, from unguided SAF in a planar substrate to near-unity coupling efficiency $\beta$ in nanowires and nanopillars. An NA\,$>$\,1 fiber guides all of the subcritical emission (green rays, $q < 1$) and part of the supercritical emission (blue rays, $1 < q \le \mathrm{NA}$), which enters the core through the near field of an emitter at $d \ll \lambda$ from the facet; rays outside its acceptance angle $\theta_\mathrm{a}$ (orange) escape into the cladding. Shading darkens with refractive index. Calculations in \textbf{a},~\textbf{b}: Lukosz--Kunz model~\cite{Lukosz1977_I,Lukosz1977_II,Lukosz1979_III}.}
    \label{fig:1}
\end{figure*}

The angular emission of a dipole at an air--dielectric interface depends on its orientation (Fig.~\ref{fig:1}a). Close to a dense medium the dipole radiates mostly into it, because the medium offers more optical states to radiate into, and the effect grows with the refractive index ($n = 1.50$ for ordinary glass, $n_{\text{core}} = 1.908$ for the fiber core used here). A dipole perpendicular to the interface (vertical, V) radiates predominantly beyond the critical angle $\theta_\mathrm{c} = \arcsin(1/n)$, whereas a parallel (horizontal, H) dipole contributes at both subcritical and supercritical angles.

The same redistribution is seen in the power radiated into the dense medium per unit normalized transverse momentum, $p(q)$ (Fig.~\ref{fig:1}b). The light line $q = 1$ separates the subcritical range, which far-field light from air can also reach, from the supercritical band $1 < q \le n$, which only the near field can feed. $p(q)$ peaks sharply at the light line and extends over the whole supercritical band. Raising the index from 1.50 to 1.908 adds emission almost exclusively at supercritical momenta.

What happens to this light depends on the structure behind the interface (Fig.~\ref{fig:1}c). In a planar substrate, SAF propagates at oblique angles without lateral confinement. A step-index fiber guides the components with $q \le \mathrm{NA}$; for a conventional fiber, this is a narrow part of the subcritical cone. A fiber with $\mathrm{NA} > 1$ moves the cutoff beyond the light line and guides the band $1 < q \le \mathrm{NA}$ in addition to all subcritical emission. Light arriving from air cannot excite these modes, because it reaches the facet only with $q \le 1$; a dipole at a subwavelength distance $d$, however, excites them through its evanescent field.

We characterize the emitter--fiber interface by the coupling efficiency
\begin{equation}
\beta = \frac{P_{\text{guided}}}{P_{\text{tot}}},
\label{eq:beta}
\end{equation}
the ratio of the guided power to the total power $P_{\text{tot}}$ radiated by the dipole into both half-spaces. For a multimode core, whose modes form a quasi-continuum in $q$, a simple estimate of $\beta$ is the momentum-acceptance fraction
\begin{equation}
\eta_{\text{acc}} = \frac{1}{P_{\text{tot}}}\int_{0}^{\mathrm{NA}} p(q) \, \mathrm{d}q ,
\label{eq:eta_acc}
\end{equation}
the fraction of the emission that enters the core at angles steep enough to be guided. We evaluate $p(q)$ and $P_{\text{tot}}$ with the plane-wave expansion of Lukosz and Kunz \cite{Lukosz1977_I, Lukosz1977_II, Lukosz1979_III, Novotny2012}, treating the core as a half-space (Methods and Supplementary Section~\ref{sec:S_model}). Fresnel transmission at the facet is included in $p(q)$; the finite core, the host nanodiamond and leaky rays are not, and are tested with full-wave simulations below. We also use $F = P_{\text{tot}}/P_0$, the total emission rate relative to that of the same dipole in a homogeneous environment (Purcell factor \cite{Purcell1946}).

\subsection*{Design landscape and distance dependence}

\begin{figure*}[!tbp]
    \centering
    \makebox[\textwidth][c]{\includegraphics[width=180mm]{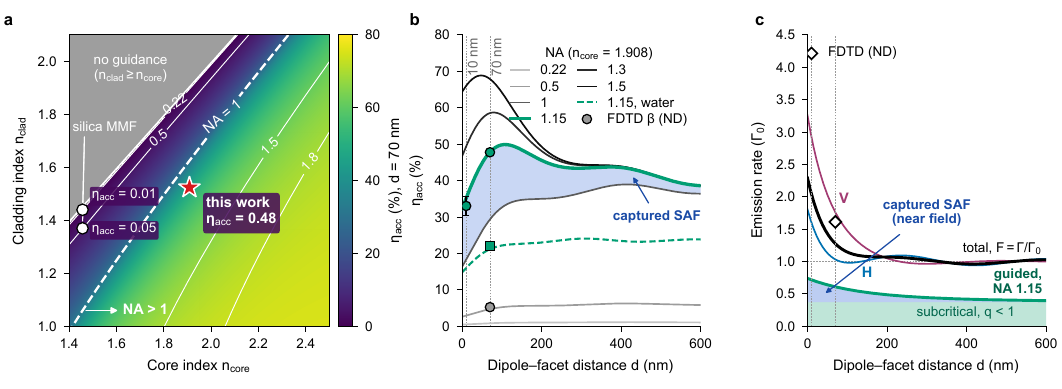}}
    \caption{\textbf{Design landscape and distance dependence of supercritical capture.}
    \textbf{a},~Orientation-averaged momentum-acceptance fraction $\eta_\mathrm{acc}$ (Eq.~\eqref{eq:eta_acc}) versus core and cladding indices, for a dipole in air \qty{70}{\nano\meter} from the facet at \qty{700}{\nano\meter}. White lines: constant NA (dashed, NA\,=\,1); grey area: no guided modes. Star: the fiber used here (NA\,=\,1.15); circles: silica fibers with NA\,=\,0.22 and 0.50.
    \textbf{b},~$\eta_\mathrm{acc}$ versus dipole--facet distance $d$ for $n_\mathrm{core} = 1.908$ and NA from 0.22 to 1.5. Thick line: this fiber; dashed: the same fiber in water. The blue band between the NA\,=\,1.15 and NA\,=\,1 curves is the captured supercritical emission. Symbols: FDTD coupling efficiency $\beta$ of an NV center in a \qty{140}{\nano\meter} nanodiamond (Fig.~\ref{fig:3}) in air (circles), in water (square) and on an NA\,=\,0.5 fiber (grey); bar at \qty{10}{\nano\meter}: range defined in Methods. Dotted lines: $d = 10$ and \qty{70}{\nano\meter}.
    \textbf{c},~Total emission rate (Purcell factor) $F = \Gamma/\Gamma_0$ of vertical (V), horizontal (H) and orientation-averaged (black) dipoles, and guided rate $\eta_\mathrm{acc}F$ of this fiber (green line), split into a subcritical part, independent of $d$ (green shading), and a supercritical part, which decays with $d$ (blue shading). Diamonds: orientation-averaged $F$ from FDTD at $d = 10$ and \qty{70}{\nano\meter}, relative to the nanodiamond without the fiber.
    Curves: Lukosz--Kunz model with the core as a half-space~\cite{Lukosz1977_I,Lukosz1977_II,Lukosz1979_III}.}
    \label{fig:2}
\end{figure*}

Figure~\ref{fig:2}a maps $\eta_{\text{acc}}$ over core and cladding indices for a dipole \qty{70}{\nano\meter} from the facet, which represents an NV center at the center of the \qty{140}{\nano\meter} NDs used here. Conventional silica fibers collect a few percent at most (5\% for $\mathrm{NA} = 0.5$), whereas our fiber ($n_{\text{core}} = 1.908$, $n_{\text{clad}} = 1.524$, $\mathrm{NA} = 1.15$ at \qty{700}{\nano\meter}) collects nearly half (48\%) of the light emitted. About 40\% of this guided light lies in the supercritical band $1 < q \le 1.15$ and would be lost by a fiber with the same core and unity NA. Because vertical and horizontal dipoles both feed this band, the capture depends only weakly on orientation, and randomly oriented NDs need no alignment.

The near field that feeds the supercritical band extends only about a tenth of a wavelength from the emitter, so the capture depends on the emitter--facet distance (Fig.~\ref{fig:2}b,c). The subcritical light entering the core does not depend on $d$. On the contrary, the supercritical part fades as the emitter moves away, as in frustrated total internal reflection. The guided emission rate $\eta_{\text{acc}}F$ therefore increases steadily as the emitter approaches the facet and roughly doubles at contact (green in Fig.~\ref{fig:2}c). The guided fraction $\eta_{\text{acc}}$ alone (without Purcell factor $F$) peaks broadly near $d \approx \qty{100}{\nano\meter}$: closer to the facet, the near field also launches light with momenta beyond the fiber cutoff ($\mathrm{NA} < q \le n_{\text{core}}$), which escapes into the cladding. For NV centers distributed throughout a \qty{140}{\nano\meter} nanodiamond, the average is close to the value at its center (Supplementary Section~\ref{sec:S_shallow}).

Because the capture relies on geometry rather than on an optical resonance, it is broadband: across the NV emission band (\qtyrange{600}{800}{\nano\meter}), the glass dispersion changes $\eta_{\text{acc}}$ only by about $\pm 5\%$ (Supplementary Section~\ref{sec:S_dispersion}). The choice of glasses balances the cutoff against material constraints: a higher $n_{\text{core}}$ widens the accessible momentum range (Fig.~\ref{fig:2}a), but dense glasses reflect more at the facet and tend to show stronger autofluorescence.

\subsection*{Full-wave simulations of dipole--fiber coupling}

\begin{figure*}[!tbp]
    \centering
    \makebox[\textwidth][c]{\includegraphics[width=180mm]{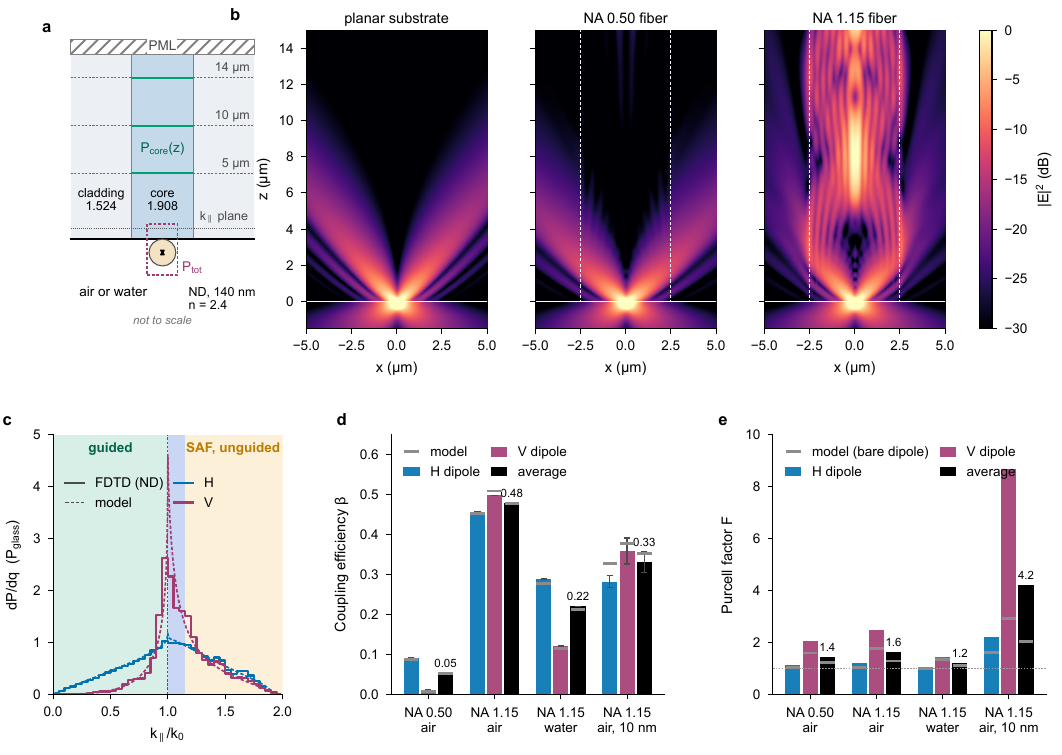}}
    \caption{\textbf{Full-wave simulation of supercritical capture.} Wavelength \qty{700}{\nano\meter}.
\textbf{a},~Geometry (not to scale): dipole in a \qty{140}{\nano\meter} nanodiamond ($n = 2.4$) on the facet of a step-index fiber (core radius \qty{2.5}{\micro\meter}) ending in an absorbing perfectly matched layer (PML). $P_\mathrm{tot}$: flux through a box around the dipole; $P_\mathrm{core}$: flux through the core 5--\qty{14}{\micro\meter} behind the facet; dotted line: plane of the momentum spectra in \textbf{c}.
\textbf{b},~Field intensity (logarithmic scale, common to the three panels) of a vertical dipole at $d = \qty{70}{\nano\meter}$ on a planar substrate ($n = 1.908$), an NA\,=\,0.5 fiber and the NA\,=\,1.15 fiber. Solid line: facet; dashed lines: core.
\textbf{c},~Power entering the glass per unit $q$, normalized to its total, for H and V dipoles at $d = \qty{70}{\nano\meter}$ on the NA\,=\,1.15 fiber: FDTD with the nanodiamond (histograms) and analytical model (dashed). Shading as in Fig.~\ref{fig:1}b.
\textbf{d},~Coupling efficiency $\beta = P_\mathrm{core}/P_\mathrm{tot}$ at $d = 70$ and \qty{10}{\nano\meter} in air or water; grey markers, analytical $\eta_\mathrm{acc}$; numbers, orientation averages; error bars, see Methods.
\textbf{e},~Purcell factor $F$ relative to the nanodiamond without the fiber; grey markers, bare dipole in air without the nanodiamond (analytical).}
    \label{fig:3}
\end{figure*}

To test the momentum-space picture beyond the half-space model, we performed three-dimensional finite-difference time-domain (FDTD) simulations of an NV center, modeled as a classical point dipole, inside a \qty{140}{\nano\meter} ND ($n = 2.4$) touching the facet of a step-index fiber (Fig.~\ref{fig:3}a and Methods). The guided power in Eq.~\eqref{eq:beta} is evaluated as the flux $P_\mathrm{core}$ through the core 5--\qty{14}{\micro\meter} behind the facet, and $F$ is taken relative to the same ND without the fiber. To limit the size of the computational domain, the simulated core has a radius of \qty{2.5}{\micro\meter}, half that of the cores of our bundle.

Figure~\ref{fig:3}b compares the emission of a vertical dipole into a planar substrate with the index of the fiber core, into an NA\,=\,0.5 fiber and into the NA\,=\,1.15 fiber. On the planar substrate and on the NA\,=\,1.15 fiber the dipole emits identically: the high-index facet decides how much light enters the glass, and the waveguide decides what happens to it afterwards. Without guiding, the supercritical emission spreads into the substrate; the NA\,=\,0.5 fiber loses almost all of it to the cladding, whereas the NA\,=\,1.15 fiber keeps a large part of it in the core.

The momentum distribution of the light entering the glass closely follows the analytical model, including the sharp maximum at the light line (Fig.~\ref{fig:3}c). For a dipole at the center of the ND, FDTD gives an orientation-averaged $\beta = 0.48$, equal to $\eta_\mathrm{acc}$, compared with 0.05 for a conventional high-NA fiber (NA\,=\,0.5), which corresponds to the bundles of Refs.~\cite{Wojciechowski2019, Jani2023} and hardly collects vertical dipoles at all (Fig.~\ref{fig:3}d): the NA\,=\,1.15 fiber guides nine times more of the emission. The ND itself changes the capture by only a few percent, because a dipole inside a subwavelength sphere radiates outside it much like a dipole at its center. The agreement persists for a shallow emitter \qty{10}{\nano\meter} from the facet and for an ND near the edge of the core, where tunneling leaky rays \cite{SnyderLove1983} even raise the guided fraction slightly (Supplementary Section~\ref{sec:S_offaxis}). The half-space estimate is therefore a reliable guide to $\beta$ under the conditions relevant here.

The capture is accompanied by a moderate, broadband increase of the emission rate (Fig.~\ref{fig:3}e). The ND strengthens the coupling of the dipole to the glass, because the space between the dipole and the core is filled with diamond ($n = 2.4$) rather than air: the emission rate increases by $F = 1.6$ on average, and up to about ninefold for a vertical dipole \qty{10}{\nano\meter} from the contact. Combining both effects, the NA\,=\,1.15 fiber delivers ten times more guided photons per unit time than the NA\,=\,0.5 fiber, without any cavity or resonance condition.

\subsection*{Design rule: $\mathrm{NA} > n_\mathrm{env}$}

The simulations indicate separate roles of two parameters: the core index relative to the environment, $n_\mathrm{core}/n_\mathrm{env}$, determines how much light enters the glass (Fig.~\ref{fig:3}b), and the numerical aperture relative to the environment, $\mathrm{NA}/n_\mathrm{env}$, determines how much of it stays guided. The light line of the environment, $q = n_\mathrm{env}$, separates waves that travel freely in the environment from the near field that only a denser medium can turn into propagating light. In air, the acceptance of the NA\,=\,1.15 fiber (Eq.~\eqref{eq:NA}) reaches past this line into the band $1 < q \le \mathrm{NA}$. In water ($n_\mathrm{env} = 1.33$) the line moves beyond the fiber NA: the acceptance window now lies entirely within light that water itself carries, the near-field advantage disappears, and the coupling efficiency drops to $\beta = 0.22$ (Fig.~\ref{fig:3}d). Near-field capture on a fiber facet therefore requires $\mathrm{NA} > n_\mathrm{env}$, independent of the emitter. For the same core glass, raising the NA above 1.33 restores the gain in water (Supplementary Section~\ref{sec:S_water} and Discussion).

\subsection*{Experimental signatures of guided high-momentum light}

\begin{figure*}[!tbp]
    \centering
    \makebox[\textwidth][c]{\includegraphics[width=180mm]{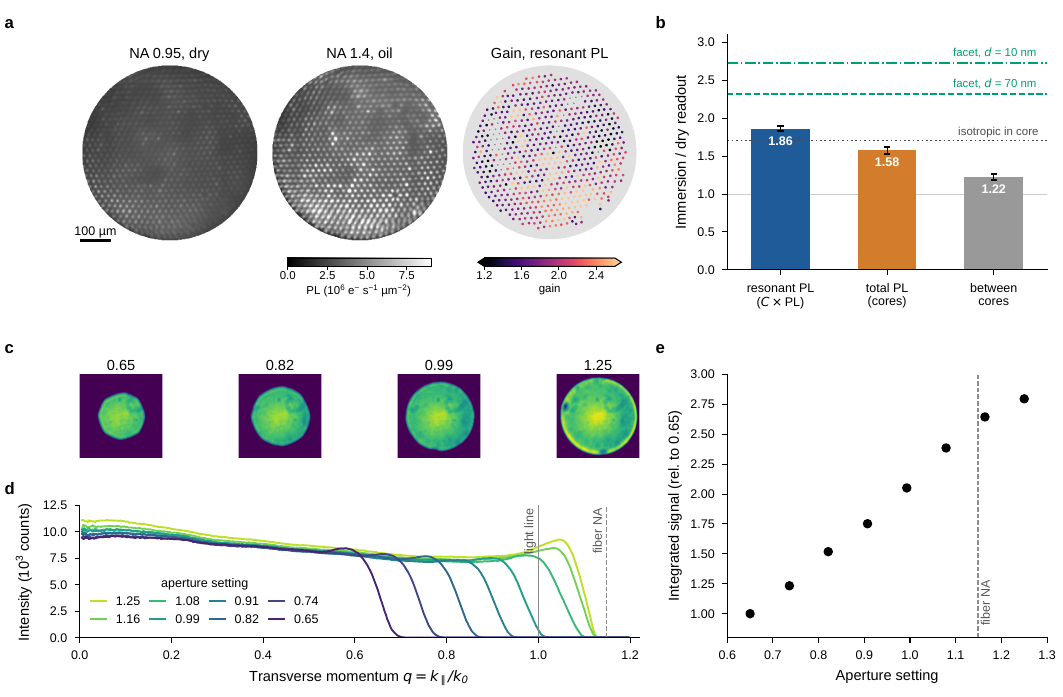}}
    \caption{\textbf{Guided high-momentum light at the output facet.}
    \textbf{a},~Output facet of the nanodiamond-coated 830-core bundle imaged through a dry ($\mathrm{NA} = 0.95$) and an oil-immersion ($\mathrm{NA} = 1.4$) $40\times$ objective at the same current of the green excitation LED: photoluminescence (PL) in photoelectrons per second and unit facet area, averaged over two measurements and the field sweep, on a common scale. Right: immersion/dry gain of the resonant PL ($C \times \mathrm{PL}$, the depth of the zero-field dip) for each core, corrected for the transmission of the objectives and the reflection of the excitation at the facet; grey dots, cores without valid fits (32 of 604). Scale bar, \qty{100}{\micro\meter}.
    \textbf{b},~Immersion/dry gain for 572 cores: resonant PL, total PL of the cores and PL between the cores (medians; error bars include the spread of the four combinations of measurements and the transmission correction). Lines: analytical model for emitters on the input facet ($d = 70$ and \qty{10}{\nano\meter}) and for isotropic emission inside the core, including the Fresnel transmission of the output facet (Supplementary Table~\ref{tab:readout}).
    \textbf{c},~Back-focal-plane (BFP) images of an oil-immersion objective with adjustable aperture, collecting the output of a second bundle (\qty{3.5}{\centi\meter} long, $\approx 800$ cores, \qty{100}{\nano\meter} nanodiamonds) excited at its input, for four of the eight aperture settings (numbers) on a color scale common to these four images.
    \textbf{d},~Azimuthally averaged BFP intensity for all eight settings versus $q$, calibrated from the pupil edges at the three smallest settings (Supplementary Fig.~\ref{fig:S_bfp_cal}). Solid and dashed lines: light line and fiber NA.
    \textbf{e},~Integrated BFP signal versus aperture setting, relative to the smallest setting (NA\,=\,0.65). Dashed line: setting equal to the fiber NA.}
    \label{fig:4}
\end{figure*}

In a straight step-index fiber without mode mixing, each guided ray keeps its transverse momentum \cite{SnyderLove1983}. Supercritical light captured at the dry input facet should therefore still carry $q > 1$ when it reaches the output facet. There it meets a dry interface beyond the critical angle and stays trapped, unless an immersion medium lets it out. We tested this prediction in two ways (Fig.~\ref{fig:4}), using two NA\,=\,1.15 imaging fiber bundles (IFBs) containing about 800 cores each, drawn from the same glasses, with input facets covalently coated with NDs: the \qty{5}{\centi\meter}-long bundle used for the endoscope below (\qty{140}{\nano\meter} NDs; Fig.~\ref{fig:4}a,b), and a \qty{3.5}{\centi\meter}-long bundle with \qty{5}{\micro\meter}-core diameter (\qty{100}{\nano\meter} NDs; Fig.~\ref{fig:4}c--e) (Methods and Supplementary Section~\ref{sec:S_setup}).

First, we imaged the output facet of the endoscope bundle through a dry ($\mathrm{NA} = 0.95$) and an oil-immersion ($\mathrm{NA} = 1.4$) $40\times$ objective (Fig.~\ref{fig:4}a). The immersion image is visibly brighter, but brightness alone is ambiguous, because the core glass also fluoresces and its light may gain differently. We therefore used the zero-field resonance (ZFR) of the NV centers, described in the next section, as a label: only light emitted by the NDs shows the resonant dip, so the depth of the dip in photoelectrons, $C \times \mathrm{PL}$ (contrast times photoluminescence), measures the ND emission alone. For 572 cores registered on the core lattice, and after correction for the transmission of the objectives and for the reflection of the excitation light at the facet (Supplementary Section~\ref{sec:S_airoil}), immersion readout increased this resonant signal per unit facet area by a factor of $1.86 \pm 0.03$ (Fig.~\ref{fig:4}b). The total photoluminescence of the cores increased less, by 1.58, and the light between the cores only by 1.22, so that the contrast of the ZFR rose by about 20\%: the additional light comes predominantly from the NDs. The measured gain lies between the analytical predictions for emitters on the input facet (2.3, including the Fresnel loss at the dry output facet) and for isotropic emission inside the core (1.7; Supplementary Table~\ref{tab:readout}), and reaches 2.1--2.2 in the brightest cores, which carry the densest ND layers. Two measurements with each objective, with opposite currents through the test structure, reproduce the gain to within 2\% and yield the same magnetic field maps (Supplementary Fig.~\ref{fig:S_airoil}).

Second, we imaged the back focal plane (BFP) of a single oil-immersion objective with an adjustable aperture (nominal NA 0.65--1.25), which collected the fluorescence at the output facet of the second bundle while the NDs were excited through a dry objective at the input facet (Fig.~\ref{fig:4}c--e and Methods). The BFP image sorts the light by direction: every radius corresponds to one value of $q$ \cite{Christinck2020, BrotonsGisbert2019}, and the excitation no longer depends on the readout aperture. We calibrated the pupil radius in units of $q$ from its linear growth with the aperture setting (Supplementary Fig.~\ref{fig:S_bfp_cal}a). With this calibration, the pupil edge follows the aperture up to settings of about 1 and then saturates at $q \approx 1.10$, close to the fiber NA: light with $1 < q \lesssim 1.1$ is transmitted by the fiber and reaches the immersion objective. The azimuthally averaged profiles (Fig.~\ref{fig:4}d) develop a maximum at $q \approx 1.03$--1.05 only when the aperture exceeds unity, where the model places the sharp maximum of near-field emission at the light line, broadened by the finite resolution of the BFP image (Supplementary Section~\ref{sec:S_bfp}); isotropic emission inside the core would produce no feature there. 
The BFP data confirm light with $q > 1$ at the output but do not reproduce the predicted angular distribution quantitatively. The maximum at the light line is weak: it rises only about 20\% above the intensity just inside the light line and stays below the intensity at the center of the pupil, whereas the model predicts about four times the central intensity for emitters on the facet. In addition, the intensity decreases by about 30\% from the center of the pupil outwards, and the integrated signal grows by a factor of 2.8 between the smallest and the largest aperture (Fig.~\ref{fig:4}e), much less than predicted for emitters on the facet (5.0--5.6) and slightly less than for isotropic emission (3.0; Supplementary Fig.~\ref{fig:S_bfp_cal}b). These deviations point to angle-dependent losses not included in the model, which attenuate high-angle light more strongly (e.g.\ bending losses of high-order modes and the transmission of the objective near its maximum aperture), and to light that fills the pupil without angular structure, such as emission redistributed by mode coupling along the bundle and, to a lesser extent, the autofluorescence of the core glass.

\subsection*{Microwave-free quantum endoscope}

\begin{figure*}[!tbp]
    \centering
    \makebox[\textwidth][c]{\includegraphics[width=180mm]{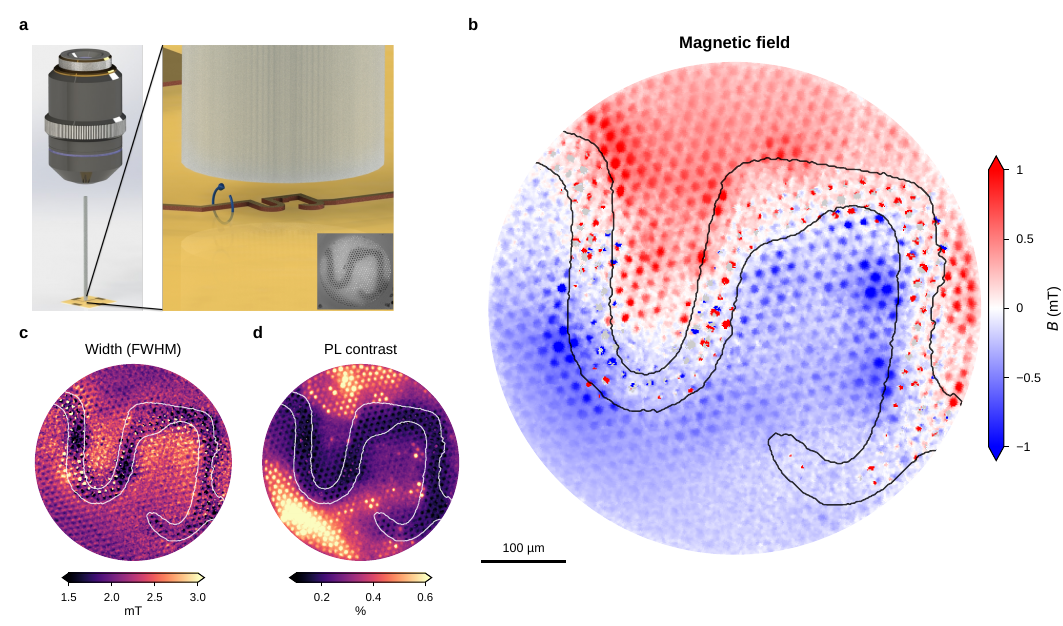}}
    \caption{\textbf{Microwave-free magnetic endoscopy through the NA\,=\,1.15 bundle.}
    \textbf{a},~Experiment: the imaging fiber bundle (IFB) between the test structure and the objective (left), and the nanodiamond-coated facet above the current-carrying `UJ' trace (right; arrow, magnetic field). Inset: the structure seen optically through the IFB.
    \textbf{b}--\textbf{d},~Parameters of the zero-field cross-relaxation resonance (ZFR) for \qty{400}{\milli\ampere} through the trace, from weighted fits of the normalized PL spectra averaged over $3 \times 3$ binned pixels. \textbf{b},~Magnetic field: ZFR position on the bias-field axis, equal in magnitude to the local field along the fiber axis (sign set by the coil orientation). \textbf{c},~Full width at half maximum of the ZFR. \textbf{d},~PL contrast of the ZFR dip.
    Readout through a dry $20\times$ objective; \qty{1.24}{\micro\meter} per binned pixel. Maps are rotated so that the lettering is upright and median-filtered over $3 \times 3$ pixels. Grey pixels in \textbf{b} (1.2\%): fits at a parameter bound or wider than the \qty{8}{\milli\tesla} sweep; values beyond $\pm\qty{1}{\milli\tesla}$ (1.3\%, mostly above the trace) are shown in the end colors; in \textbf{c},\,\textbf{d} the 0.7\% of failed fits are replaced by the median of their neighbours. Lines: outline of the trace from a confocal image; the lower arc of the `J' carries no current. Scale bar (\textbf{b}), \qty{100}{\micro\meter}; \textbf{c} and \textbf{d} show the same field of view, the whole bundle (diameter \qty{578}{\micro\meter}).}
    \label{fig:endoscope}
\end{figure*}

We used the ND-coated bundle for lensless endoscopic magnetic-field imaging (Fig.~\ref{fig:endoscope}a). The coated facet was placed above a test structure---a copper trace about \qty{50}{\micro\meter} wide (designed width) shaped as the letters `UJ' on a flexible printed circuit board---while the opposite facet was imaged with a wide-field fluorescence microscope through a dry $20\times$ objective ($\mathrm{NA} = 0.8$), which also delivered the green LED excitation light. This lower magnification images the whole bundle onto the camera, but because the readout is dry, the maps use only the subcritical part of the guided light.

The magnetic field was read out without microwaves, using the ZFR, a near-zero-field cross-relaxation feature of NV ensembles \cite{Dhungel2024, Sengottuvel2025, Dhungel2026}. At zero field, NV centers of different orientations, and nearby paramagnetic defects, come into resonance with one another and exchange spin polarization (cross-relaxation); their spin relaxation speeds up and the photoluminescence (PL) dims. Any local field detunes them, so the dip appears at the bias field that cancels the local field. We swept the bias field along the fiber axis and recorded, at every field step, an image of the PL together with a reference image with the bias coil switched off, which removes slow drifts of the excitation and makes the field map insensitive to variations of the ND coverage. At each binned pixel, the normalized PL spectrum was fitted to obtain the position of the resonance, its full width at half maximum (FWHM) $w$ and its PL contrast $C$ (Methods).

Figure~\ref{fig:endoscope}b shows the field map for a direct current of \qty{400}{\milli\ampere} through the trace, updating a preliminary map from the same dataset~\cite{Wojciechowski2025SPIE} with image-subtraction overflow errors fixed and complemented by maps of the resonance width and contrast (Fig.~\ref{fig:endoscope}c,d). The resonance position reproduces the geometry of the conductor, with shifts of opposite sign on the two sides of the trace, as expected for the component of the Oersted field along the fiber axis. A Biot--Savart calculation for the current path extracted from a confocal image of the structure reproduces the map without a free amplitude, for a distance of about \qty{100}{\micro\meter} between the conductor and the NDs (Supplementary Fig.~\ref{fig:S_oersted}). The bundle was held above the structure without contact, leaving a gap of this order. This standoff limits the field at the facet to about $\pm\qty{0.5}{\milli\tesla}$ and blurs it over a comparable distance.

The conductor is seen most clearly in the contrast map (Fig.~\ref{fig:endoscope}d): above the trace the contrast falls from 0.35\% to about 0.2\%. Directly above a current the Oersted field lies almost parallel to the facet, whereas the bias coil sweeps only the axial component. Above the trace the total field therefore never passes through zero: the sweep can cancel the axial field but leaves the transverse one, so the cross-relaxation between differently oriented NV centers remains incomplete and the dip is shallower. The Biot--Savart model places transverse fields of up to \qty{0.6}{\milli\tesla} exactly where the contrast drops, whereas the gradient of the axial field across a single core (below \qty{0.1}{\milli\tesla}) is far too small to broaden the resonance. Accordingly, the fitted width (Fig.~\ref{fig:endoscope}c) does not increase on average above the trace, but scatters more from core to core, as expected for a low-contrast resonance fitted with a single Gaussian. The sampling is set by the core pitch of \qty{19}{\micro\meter}, the resolution of the field map here by the standoff.

The sensitivity follows from two measured quantities (Supplementary Section~\ref{sec:S_sensitivity}): the steepest slope of the resonance, given by its fitted contrast and width, and the noise of the normalized signal, obtained from the scatter between the 16 repetitions of the sweep. The noise equals the photon shot noise to within 20\%. The resulting sensitivity is that of a measurement at the working point on the flank of the resonance, where a small change of the field produces the largest change of the signal: $\eta_B \approx \qty{0.23}{\milli\tesla\per\sqrt{\hertz}}$ for one cell of the core lattice (\qty{300}{\micro\meter\squared}) over the measurement time, or \qty{0.06}{\milli\tesla\per\sqrt{\hertz}} if only the camera exposure is counted. The area-normalized sensitivity, $\eta_B\sqrt{A} \approx \qty{4}{\milli\tesla\micro\meter\per\sqrt{\hertz}}$ (1.1 for the exposure), allows comparison with other imagers \cite{Levine2019}. The maps themselves, which sample the whole resonance, determine the field away from the trace to \qty{15}{\micro\tesla} per cell in \qty{42}{\minute}. These values hold for each of about 800 cells recorded simultaneously: a single-point fiber probe with the same per-point sensitivity would need about 800 times longer to map the same area by scanning. The NDs set the contrast and width of the resonance, the fiber sets the photon budget. Since the shot-noise-limited sensitivity scales as $w/(C\sqrt{I_0})$, with $I_0$ the detected photoelectron rate, the gain from $\mathrm{NA} > 1$ enters through $I_0$.

The same measurement with the $40\times$ objectives of Fig.~\ref{fig:4}a, at one third of the excitation current, was 1.5 (dry) and 2.2 times (immersion) more sensitive per unit exposure time than the $20\times$ map (Supplementary Table~\ref{tab:freespace}), but its maps are noisier, because the total exposure was 7 to 14 times shorter. For comparison, the same protocol and analysis applied to a layer of the same NDs imaged directly through a dry $40\times$/0.65 objective at the LED current of Fig.~\ref{fig:endoscope} \cite{Sengottuvel2025} give a contrast of 0.7--0.8\%, a width of \qty{2.4}{\milli\tesla} and an exposure-limited area-normalized sensitivity of 0.20--\qty{0.23}{\milli\tesla\micro\meter\per\sqrt{\hertz}}. Immersion readout through the bundle reaches \qty{0.53}{\milli\tesla\micro\meter\per\sqrt{\hertz}} at one third of that current, or about 0.3 if the signal scales linearly with the excitation (Supplementary Table~\ref{tab:freespace}). Read out through \qty{5}{\centi\meter} of the bundle, the NDs on the facet thus deliver field information per unit area within a factor of about 1.5 of NDs imaged directly, despite the lower contrast caused by the fiber background. Wide-field microscopes based on dense NV layers in single-crystal diamond reach area-normalized sensitivities orders of magnitude better, but require a planar diamond sample at the object; for nanodiamond sensors, the relevant benchmark is the same NDs imaged directly.

\section*{Discussion}

We have shown that optical fibers with $\mathrm{NA} > 1$ (Eq.~\eqref{eq:NA}) act as a near-field collection interface between nanoscale emitters and macroscopic multicore fibers. The mechanism is momentum matching rather than an optical resonance. The core index relative to the environment sets how much of the near field of the dipole is drawn into the glass, and the condition $\mathrm{NA} > n_\mathrm{env}$ decides whether its supercritical part is guided or lost to the cladding. The analytical model and the FDTD simulations agree on a coupling efficiency of 0.48 for an NA\,=\,1.15 fiber, nine times that of an NA\,=\,0.5 fiber, with about 40\% of the guided light in the supercritical band, a broadband emission-rate enhancement of $F \approx 1.6$, and nearly uniform capture across the NV emission band.

Both experiments of Fig.~\ref{fig:4} support the one-way-door picture of the Introduction, in which dry near-field capture at the sample and immersion readout at the remote end are complementary rather than contradictory: light with $q > 1$ arrives at the output facet, and an immersion objective extracts 1.9 times more of the ND emission than a dry one. In the magnetic measurement this gain improves the sensitivity 1.4-fold, which halves the measurement time for a given precision; it acts mainly through a higher contrast, because the additional light is ND emission rather than background. In the endoscope, the resonance position reproduces the Oersted field of a current-carrying conductor without a free amplitude, and the maps of resonance width and contrast separate two effects of this field: its transverse component reduces the depth of the resonance, whereas its gradient across a core is too small to broaden it.

Nanostructures built around single emitters (Introduction) reach $\beta > 0.9$ but require nanofabrication or deterministic positioning, and some rely on resonant, narrowband Purcell enhancement \cite{Purcell1946, Joe2026}. All-diamond scanning probes \cite{Maletinsky2012, Appel2016, Hedrich2020} offer nanometer-scale resolution at the cost of scanning a single sensor. Conventional multimode fibers are scalable but capture at most about 5\% of the dipole emission. NA\,$>$\,1 fibers lie in between: they accept randomly oriented NDs attached by simple chemistry, and come as bundles of hundreds of parallel cores, trading spatial resolution for a parallel, wide-field measurement through a passive and robust element.

Several limitations should be noted. First, the experimental evidence for supercritical capture is still indirect: neither the immersion gain nor the back-focal-plane profiles separate near-field capture from high-angle scattering by the NDs or from mode mixing. Dielectric spacers that vary the emitter--facet distance, and measurements of the predicted lifetime shortening, would provide direct tests. Second, the supercritical gain is restricted to dry operation; in water the present fiber collects $\beta = 0.22$, whereas near-field capture requires $\mathrm{NA} > 1.33$ (see below). Third, the cores occupy about a quarter of the facet, so only 11--12\% of the emission of all NDs on the facet is coupled into core modes, while emitters above the cladding contribute background (Supplementary Section~\ref{sec:S_offaxis}). Fourth, extracting the supercritical modes requires immersion optics at the readout end; such objectives have high magnification, so imaging the whole bundle requires a short tube lens or a large camera sensor. Fifth, and most important for the magnetic measurements, the sensitivity reported here is limited by the camera rather than by the fiber: with immersion readout the bundle delivered nearly twice as many photoelectrons as the camera pixels could store, and the sequential field sweep exposed the camera for only 1--7\% of the time (Supplementary Table~\ref{tab:camera}). Cameras with deeper wells and continuous acquisition \cite{Wojciechowski2018} should improve the sensitivity by an order of magnitude or more without any change to the fiber.

The design rules of Fig.~\ref{fig:2}a show how these limits can be relaxed. Keeping the present core glass and replacing the glass cladding with a low-index material raises the numerical aperture above the index of water: $\mathrm{NA} > 1.33$ requires $n_\mathrm{clad} < 1.365$, which amorphous fluoropolymers such as Teflon AF and CYTOP and UV-curable cladding polymers meet ($n \approx 1.31$--1.34) \cite{SpecialChemLowRI}; with the present core glass they give $\mathrm{NA} = 1.36$--1.39 and up to twice the capture in water. A suspended core, held by thin bridges inside air holes, offers an even larger contrast, provided the holes are sealed at the facet so that water cannot fill them. Drawing multi-core bundles with polymer claddings or suspended cores remains a fabrication challenge. Air-clad polymer fibers are another route: fibers with $\mathrm{NA} \approx 1$ have recently become commercially available \cite{SHUTE2026}, and a poly(methyl methacrylate) core ($n \approx 1.49$) with a cladding index close to 1 reaches $\mathrm{NA} \approx 1.1$, for which the model gives $\eta_\mathrm{acc} = 0.50$ ($d = \qty{70}{\nano\meter}$), nine times more than an NA\,=\,0.5 fiber, in a flexible format, although its NA remains below 1.33 in water. The sampling, set here by the \qty{19}{\micro\meter} core pitch, is not a fundamental limit either: the same glass pair has been drawn into bundles with cores down to 1.3--\qty{1.5}{\micro\meter} at a pitch of about \qty{2}{\micro\meter} (Supplementary Table~\ref{tab:bundles}; see also Ref.~\cite{Bijoch2023}), and ND-coated bundles of other glasses with $\mathrm{NA} \approx 0.5$ and about 15\,000 cores have already served as magnetic imagers \cite{Wojciechowski2019, Jani2023}. Nothing prevents a similar pixel count with an $\mathrm{NA} > 1$ glass pair, whose higher index contrast also confines the light more tightly in closely spaced cores. A \qty{2.7}{\micro\meter} core still supports about a hundred modes at $\mathrm{NA} = 1.15$, enough for the multimode estimate of the coupling efficiency; for the smallest cores, with a few tens of modes, it becomes approximate.

The endoscope demonstrated here combines dry near-field capture at the sample with all-optical, microwave-free magnetometry, and its signal increases with immersion readout at the remote end. This combination is attractive where microwave delivery and bulk immersion optics are impractical, for example for imaging superconductors inside cryostats \cite{Dhungel2026}. It also brings SAF detection, which so far requires an immersion objective at the sample, to a passive fiber facet. The same interface is of interest for single emitters---molecules, fluorescent proteins, quantum dots and color centers \cite{Brooks1990, Aharonovich2016}---whose signal is limited by saturation or photobleaching. In the shot-noise limit, doubling the collection efficiency halves the measurement time for a given signal-to-noise ratio, or doubles the number of photons detected before an emitter bleaches, and each core of a bundle can serve as an independent detection channel. Because the capture is set by the numerical aperture rather than by a particular glass, core and cladding glasses can also be chosen for lower autofluorescence than the present lead--bismuth--gallate glass. Once fibers with $\mathrm{NA} > 1.33$ retain the near-field gain in aqueous media, the same architecture would become available for biological applications of fluorescent probes and ND sensors \cite{Kolodziej2024}, where halving the acquisition time or the excitation dose can decide whether fast or light-sensitive processes in living samples can be followed at all. More generally, for a fiber facing a nanoscale emitter the relevant comparison is not between NA and 1 but between NA and $n_\mathrm{env}$; the momentum-space design rules given here apply to any emitter within a fraction of a wavelength of a high-contrast fiber facet.

\section*{Methods}

\subsection*{Analytical model}

The fiber core was treated as a dielectric half-space of index $n_\mathrm{core}$ and the emitter as an electric point dipole at a distance $d$ from the interface, in an environment of index $n_\mathrm{env}$ (air or water). The emitted field was expanded in plane waves following Lukosz and Kunz \cite{Lukosz1977_I, Lukosz1977_II, Lukosz1979_III} and Ref.~\cite{Novotny2012}; the expressions for the power transmitted into the core per unit transverse momentum, $p(q)$, and for the total emitted power $P_\mathrm{tot}$ are given in Supplementary Section~\ref{sec:S_model}. The momentum-acceptance fraction $\eta_\mathrm{acc}$ (Eq.~\eqref{eq:eta_acc}) is the integral of $p(q)$ up to $q = \mathrm{NA}$ divided by $P_\mathrm{tot}$. Orientation averages were calculated as $(P_\mathrm{V} + 2P_\mathrm{H})/3$ for both the guided and the total power. The implementation was validated against the closed-form expressions for $d = 0$, against direct numerical integration of the total power for $d > 0$, and against the free-space limit; the code is available in the data repository \cite{dataset_UJ}.

\subsection*{Finite-difference time-domain simulations}

Simulations were performed with the FDTD method \cite{Taflove2005} using the GPU-based solver Tidy3D (version 2.11.2)~\cite{Tidy3D}. The fiber was modeled as a semi-infinite step-index waveguide with a core radius of \qty{2.5}{\micro\meter}, extending from the facet through the perfectly matched layer (PML) at the far end of the simulation domain. The refractive indices of the NA\,$>$\,1 fiber were $n_\mathrm{core} = 1.908$ and $n_\mathrm{clad} = 1.524$ (Sellmeier values at \qty{700}{\nano\meter}, NA\,=\,1.149); the low-NA reference fiber had $n_\mathrm{core} = 1.60$ and $n_\mathrm{clad} = 1.52$ (NA\,=\,0.50). The ND was a sphere of \qty{140}{\nano\meter} diameter with $n = 2.4$ touching the facet, surrounded by air or water ($n = 1.333$). The NV center was a point dipole at the center of the ND ($d = \qty{70}{\nano\meter}$) or \qty{10}{\nano\meter} from the facet, driven by a Gaussian pulse centered at \qty{700}{\nano\meter} with a bandwidth of $f_0/10$. The simulation domain was $10 \times 10 \times 16.5$\,\unit{\micro\meter\cubed} with PML on all sides. The automatic mesh (20 cells per wavelength in the material) was refined to \qty{8}{\nano\meter} around the ND, and to \qty{3}{\nano\meter} for $d = \qty{10}{\nano\meter}$; refining further to \qty{2}{\nano\meter} changed the Purcell factor and $\beta$ by less than 0.1\%. All simulations were run until the fields decayed below $10^{-6}$ of their maximum.

The total emitted power $P_\mathrm{tot}$ was obtained from the flux through a closed box (\qty{0.4}{\micro\meter}) around the dipole, and the guided power from the flux through the core cross-section at $z = 5$ and \qty{10}{\micro\meter} behind the facet, and in the later simulations ($d = \qty{10}{\nano\meter}$, the off-axis emitter and the planar substrate) also at \qty{14}{\micro\meter}; $\beta$ is given for $z = \qty{10}{\micro\meter}$, where it had converged to within 1.5\% for on-axis emitters. For $d = \qty{10}{\nano\meter}$, where light beyond the cutoff leaves the core region slowly, we report the range between the flux through the circular core ($z = \qty{12}{\micro\meter}$) and through a square enclosing the core ($z = \qty{14}{\micro\meter}$). The Purcell factor was defined as $P_\mathrm{tot}$ divided by the power of the same ND-hosted dipole in the homogeneous environment without the fiber, computed with the identical local mesh. For on-axis dipoles the $x$ and $y$ orientations are equivalent by symmetry; orientation averages were calculated as $(2P_\mathrm{H} + P_\mathrm{V})/3$ for both the guided and the total power. The momentum spectra in Fig.~\ref{fig:3}c were obtained by Fourier transforming the tangential fields on a plane \qty{0.3}{\micro\meter} behind the facet and computing the axial Poynting flux per unit in-plane wavevector. An overview of all simulations and convergence tests is given in Supplementary Section~\ref{sec:S_fdtd}.

\subsection*{Fiber fabrication and nanodiamond functionalization}

The imaging fiber bundles were fabricated from an in-house-synthesized lead--bismuth--gallate heavy-metal-oxide glass (PBG81) for the cores and a lower-index borosilicate-type glass (UV710) for the cladding, in a structure similar to that of Ref.~\cite{Bijoch2023}. At \qty{700}{\nano\meter} the glasses have $n_{\text{core}} = 1.908$ and $n_{\text{clad}} = 1.524$, giving $\mathrm{NA} = 1.15$ (Supplementary Section~\ref{sec:S_dispersion}); the numerical aperture was verified independently with an immersion goniometer at \qty{638}{\nano\meter} and \qty{975}{\nano\meter} (Supplementary Section~\ref{sec:S_NA}). Two rigid bundles made of these glasses were used. The endoscope bundle has an imaging diameter of \qty{578}{\micro\meter}, measured with a confocal microscope (Olympus FV3000, $10\times$ objective), and $\approx 830$ cores with a diameter of \qty{10}{\micro\meter} and a pitch of \qty{19.2}{\micro\meter} (nominally \qty{20}{\micro\meter}; from the calibrated confocal image, consistent with the number of cores in the imaging area and with the magnification of the readout); the number of cores was obtained by fitting a hexagonal lattice to images of the facet and classifying each lattice site (820--831 depending on the classification of the outermost row). The bundle used for back-focal-plane imaging has an imaging diameter of about \qty{267}{\micro\meter} (outer diameter \qty{567}{\micro\meter}) and $\approx 800$ cores with a diameter of 4.8--\qty{5.2}{\micro\meter} and a pitch of 8.7--\qty{9.1}{\micro\meter}.

Carboxylated NDs with a nominal diameter of \qty{140}{\nano\meter} (Adamas Nanotechnologies; $\sim 3$~ppm NV, corresponding to $\approx 760$ NV centers per particle) were covalently attached to the polished facet of the \qty{5}{\centi\meter}-long endoscope bundle by silanization with (3-aminopropyl)triethoxysilane (APTES) followed by 1-ethyl-3-(3-dimethylaminopropyl)carbodiimide (EDC) cross-linking \cite{Czarnecka2022, Jani2023}. The process yielded submonolayer coverage with an estimated density of $\sim 15$ NDs per \unit{\micro\meter\squared} (approximately 23\% of the area). The \qty{3.5}{\centi\meter}-long bundle used for back-focal-plane imaging was coated in the same way with \qty{100}{\nano\meter} NDs (Adamas Nanotechnologies).

\subsection*{Optical characterization}

In the back-focal-plane measurements, fluorescence was excited with a continuous-wave \qty{532}{\nano\meter} laser; in all wide-field images of the output facet (Figs.~\ref{fig:4}a,b and \ref{fig:endoscope}), the green LED described under Magnetic-field imaging was focused onto the back focal plane of the readout objective. For back-focal-plane imaging (Supplementary Fig.~\ref{fig:Fiber-setup}b), \qty{1}{\milli\watt} of excitation light was focused onto the input facet of the \qty{3.5}{\centi\meter}-long bundle with a dry objective facing the coated facet from the air side ($50\times$, $\mathrm{NA} = 0.8$, Olympus MPLFLN50X), and the fluorescence at the output facet was collected with a $60\times$ oil-immersion objective with a variable aperture ($\mathrm{NA} = 0.65$--1.25, Olympus UPLFLN 60XOI; immersion oil $n = 1.518$). After a long-pass filter (Thorlabs FELH0600), a Bertrand lens ($f = \qty{250}{\milli\meter}$) imaged the objective pupil onto the camera (IDS U3-3240CP). The aperture was set to eight equally spaced nominal values between 0.65 and 1.25. The pupil radius was calibrated in units of $q$ from the pupil edges at the three smallest settings (\qty{388}{px} per unit NA; Supplementary Section~\ref{sec:S_bfp}).

\subsection*{Magnetic-field imaging}

The NDs were excited by a green LED (Thorlabs M530L4; peak wavelength \qty{522}{\nano\meter}, FWHM \qty{35}{\nano\meter}, \qty{480}{\milli\watt} at the maximum current of \qty{1}{\ampere}) focused onto the back focal plane of the readout objective (Olympus UPLXAPO20X, $20\times$, $\mathrm{NA} = 0.8$), which illuminated the facet of the bundle uniformly. The LED current was \qty{0.9}{\ampere} (maximum \qty{1}{\ampere}); the power at the facet was not measured. The coated facet was positioned above the test structure, without contact, with a manual translation stage. The red photoluminescence returning through the bundle was collected by the same objective, separated by a dichroic mirror, filtered with a \qty{600}{\nano\meter} long-pass filter (Thorlabs FELH0600), and imaged with a \qty{100}{\milli\meter} tube lens onto a 12-bit CMOS camera (Basler acA2440-35um, Sony IMX264 sensor, $2448 \times 2048$ pixels of \qty{3.45}{\micro\meter}). A bias magnetic field along the fiber axis was applied with a custom three-axis Helmholtz coil system and stepped from $-4.0$ to \qty{3.9}{\milli\tesla} in steps of \qty{0.1}{\milli\tesla}. At every step, one image was recorded with the bias field on and one reference image with the bias coil current switched off (exposure \qty{70}{\milli\second} each); the sweep was repeated 16 times, giving a total measurement time of \qty{42.5}{\minute}. After $4 \times 4$ binning, the ratio of the summed reference and signal images was calculated at each binned pixel and field step, with its standard error obtained from the scatter between the 16 repetitions. The spectra were averaged over $3 \times 3$ binned pixels and fitted by weighted least squares with a Gaussian on a constant baseline, giving the position, FWHM and relative height $\varepsilon$ of the cross-relaxation resonance \cite{Sengottuvel2025, Dhungel2026} (median reduced $\chi^2 = 1.26$); the relative height corresponds to a PL contrast $C = \varepsilon/(1 + \varepsilon)$. Camera pixels at or above \qty{4000}{\ADU} (0.03\%) were excluded. The spatial scale of the maps, \qty{1.24}{\micro\meter} per binned pixel, follows from the magnification ($f = \qty{9}{\milli\meter}$ objective and \qty{100}{\milli\meter} tube lens) and agrees to within 1\% with the imaging diameter measured with the confocal microscope. The Oersted field of the trace was calculated with the Biot--Savart law for the current path extracted from a confocal image of the structure (Supplementary Section~\ref{sec:S_oersted}). For comparison of dry and immersion readout, the $20\times$ objective was replaced by $40\times$ objectives with $\mathrm{NA} = 0.95$ (dry, Olympus UPLXAPO40X) and $\mathrm{NA} = 1.4$ (oil immersion, Olympus UPLXAPO40XO), and the tube lens was replaced by a $f = \qty{50}{\milli\meter}$ lens mounted in SM1 lens tubes (Thorlabs), so the whole bundle was imaged. Because the distance between this lens and the camera sensor did not equal its focal length, the scale of these maps, \qty{1.46}{\micro\meter} per binned pixel (effective focal length $\approx \qty{42}{\milli\meter}$), was calibrated from the core pitch of the bundle (\qty{19.2}{\micro\meter}) rather than taken from the nominal magnification. These maps were recorded with an LED current of \qty{0.3}{\ampere}, exposures of \qty{10}{\milli\second} (dry) and \qty{5}{\milli\second} (oil), 16 sweeps each, and for both directions of the current through the trace (Supplementary Section~\ref{sec:S_airoil}). All measurements were performed at room temperature.

\subsection*{Use of AI tools}

Generative AI tools were used under the supervision of the authors. Claude (Anthropic) assisted in writing and checking the code for the analytical model, the FDTD simulations and their analysis, the processing and fitting of the camera data, the image registration and the figures, and in revising the text and captions. Gemini (Google) was used for early-version language editing, LaTeX formatting, and reference organization. All code was validated against analytical limits, power conservation, and independent simulations. The authors reviewed all text, figures, and conclusions and assume full responsibility for the content. No AI-generated images were used.

\backmatter

\bmhead{Data availability}
The data underlying the results presented in this paper, as well as the simulation codes and analysis scripts, are publicly available at Ref.~\cite{dataset_UJ}.

\bmhead{Acknowledgements}
The authors thank Wojciech Gawlik for inspiring discussions, and Paulina Czarnecka and Mona Jani for preparing the nanodiamond-coated optical fibers. The first batch of NA\,$>$\,1 fibers was developed within the TEAM-NET programme of the Foundation for Polish Science co-financed by the European Union
under the European Regional Development Fund, project POIR.04.04.00-00-1644/18. This work was supported by the National Science Centre, Poland, grant no.~2023/05/Y/ST3/00135 (AQuSeND) within the QuantERA II Programme, which has received funding from the European Union's Horizon 2020 research and innovation programme under Grant Agreement No.~101017733. The study was carried out using research infrastructure purchased with the funds of the European Union in the framework of the Smart Growth Operational Programme, Measure~4.2; Grant No.~POIR.04.02.00-00-D001/20, ``ATOMIN~2.0 -- ATOMic scale science for the INnovative economy''. For the purpose of Open Access, the author has applied a CC-BY public copyright license to any Author Accepted Manuscript (AAM) version arising from this submission.

\section*{Declarations}

\textbf{Author contributions}\\
M.M. built the magnetic-endoscopy setup and performed the magnetic-field imaging experiments, performed the back-focal-plane experiments, contributed to the data analysis and drafted early versions of the manuscript. S.S. wrote the code for map acquisition and analysis, performed the FDTD simulations and contributed to the data analysis. A.F., G.S., D.P. and R.B. developed and fabricated the imaging fiber bundles and characterized them (optical, SEM and NA measurements). A.M.W. conceived and supervised the study, analysed the data, acquired funding and wrote the manuscript. All authors reviewed and edited the manuscript.
\\
\textbf{Competing interests}\\
A.M.W. is an author of a related patent application. All other authors declare no competing interests.

\bibliography{bib.bib}

\newpage
\begin{appendices}
\setcounter{figure}{0}
\setcounter{table}{0}
\renewcommand{\thefigure}{A\arabic{figure}}
\renewcommand{\thetable}{A\arabic{table}}

\section{Supplementary Information}\label{secA1}

\subsection{Analytical model of dipole emission at the fiber facet}
\label{sec:S_model}

The fiber core is treated as a dielectric half-space of index $n_\mathrm{core}$, and the emitter as an electric point dipole in the environment (index $n_\mathrm{env}$, air or water) at a distance $d$ from the facet. The emitted field is expanded in plane waves with in-plane wavevector $k_\parallel = s\,n_\mathrm{env}k_0$, so that $s = q/n_\mathrm{env}$ with $q$ of the main text, following Lukosz and Kunz~\cite{Lukosz1977_I,Lukosz1977_II,Lukosz1979_III} and Ref.~\cite{Novotny2012}. With $m = n_\mathrm{core}/n_\mathrm{env}$, $k_{z1} = \sqrt{1-s^2}$, $k_{z2} = \sqrt{m^2-s^2}$ and the Fresnel coefficients
\begin{equation}
r_\mathrm{p} = \frac{m^2k_{z1} - k_{z2}}{m^2k_{z1} + k_{z2}}, \qquad
r_\mathrm{s} = \frac{k_{z1} - k_{z2}}{k_{z1} + k_{z2}},
\end{equation}
the power transmitted into the core per unit $s$, in units of the power $P_0$ of the same dipole in the homogeneous environment, is for $s<1$
\begin{align}
\frac{\mathrm{d}P_\mathrm{V}}{\mathrm{d}s} &= \frac{3}{4}\,\frac{s^3}{k_{z1}}\left(1-|r_\mathrm{p}|^2\right), &
\frac{\mathrm{d}P_\mathrm{H}}{\mathrm{d}s} &= \frac{3}{8}\,s\left[\frac{1-|r_\mathrm{s}|^2}{k_{z1}} + k_{z1}\left(1-|r_\mathrm{p}|^2\right)\right],
\end{align}
and for the supercritical range $1 < s < m$, where $k_{z1} = \mathrm{i}\sqrt{s^2-1}$ is imaginary,
\begin{align}
\frac{\mathrm{d}P_\mathrm{V}}{\mathrm{d}s} &= \frac{3}{2}\,\mathrm{Re}\!\left[\frac{s^3}{k_{z1}}\,r_\mathrm{p}\,e^{2\mathrm{i}n_\mathrm{env}k_0dk_{z1}}\right], &
\frac{\mathrm{d}P_\mathrm{H}}{\mathrm{d}s} &= \frac{3}{4}\,\mathrm{Re}\!\left[s\left(\frac{r_\mathrm{s}}{k_{z1}} - k_{z1}r_\mathrm{p}\right)e^{2\mathrm{i}n_\mathrm{env}k_0dk_{z1}}\right],
\end{align}
where H denotes the azimuthal average for a horizontal dipole. The supercritical contribution decays with $d$ through the evanescent factor, whereas the subcritical one does not depend on $d$. The total emitted power (Purcell factor) $F = P_\mathrm{tot}/P_0$ is obtained from the corresponding expressions for the total power~\cite{Novotny2012}, including emission into the environment, and the momentum-acceptance fraction of a fiber with numerical aperture NA is
\begin{equation}
\eta_\mathrm{acc} = \frac{1}{F}\int_0^{\mathrm{NA}/n_\mathrm{env}}\frac{\mathrm{d}P}{\mathrm{d}s}\,\mathrm{d}s,
\end{equation}
which is Eq.~\eqref{eq:eta_acc} of the main text. Orientation averages are $(P_\mathrm{V} + 2P_\mathrm{H})/3$ for both the guided and the total power.

The implementation was validated against the closed-form Lukosz--Kunz expressions for $d = 0$ (relative deviation $<10^{-8}$), against direct numerical integration of the total-power integrals for $d>0$ (relative deviation $<10^{-6}$) and against the free-space limit $n_\mathrm{core}\to n_\mathrm{env}$. The code is included in the data repository~\cite{dataset_UJ}.

\subsection{Refractive-index dispersion and spectral dependence of the capture}
\label{sec:S_dispersion}

\begin{figure*}[ht]
 \centering
 \includegraphics[width=0.85\textwidth]{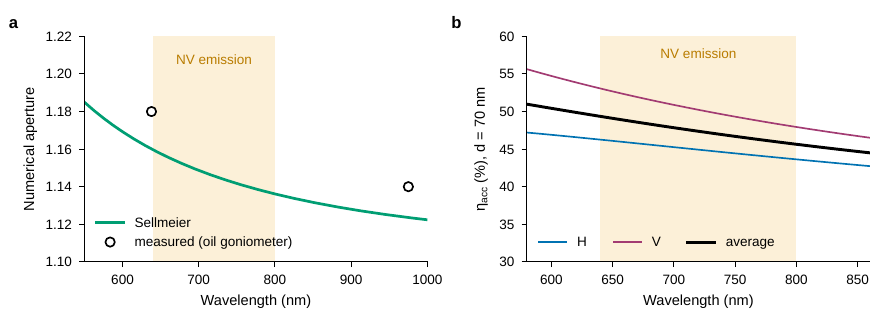}
 \caption{\textbf{Spectral dependence.} \textbf{a},~Numerical aperture calculated from the Sellmeier coefficients (line) and measured with the immersion goniometer (symbols, Section~\ref{sec:S_NA}). \textbf{b},~Momentum-acceptance fraction $\eta_\mathrm{acc}$ ($d = \qty{70}{\nano\meter}$) for horizontal (H), vertical (V) and orientation-averaged dipoles, calculated with the dispersive core and cladding indices. Shaded: NV emission band.}
 \label{fig:S_dispersion}
\end{figure*}

The refractive indices of the core (PBG81) and cladding (UV710) glasses were described by the Sellmeier equation
\begin{equation}
n^2(\lambda) = 1 + \sum_{i=1}^{3}\frac{B_i\lambda^2}{\lambda^2 - C_i},
\end{equation}
with the coefficients listed in Table~\ref{tab:sellmeier} ($\lambda$ in \unit{\micro\meter}). At \qty{700}{\nano\meter} this gives $n_\mathrm{core} = 1.908$, $n_\mathrm{clad} = 1.524$ and $\mathrm{NA} = 1.149$. Across the NV emission band the numerical aperture decreases from 1.169 at \qty{600}{\nano\meter} to 1.136 at \qty{800}{\nano\meter}, and $\eta_\mathrm{acc}$ ($d = \qty{70}{\nano\meter}$, orientation averaged) from 0.504 to 0.456 (Fig.~\ref{fig:S_dispersion}). The capture is therefore broadband, varying by only $\pm5\%$ over the phonon sideband.

\begin{table}[ht]
\centering
\caption{Sellmeier coefficients of the core and cladding glasses ($C_i$ in \unit{\micro\meter\squared}).}
\label{tab:sellmeier}
\begin{tabular}{lcccccc}
\hline
Glass & $B_1$ & $B_2$ & $B_3$ & $C_1$ & $C_2$ & $C_3$ \\
\hline
PBG81 (core) & 2.033040 & 0.474500 & 0.299980 & 0.015300 & 0.065120 & 41.542800 \\
UV710 (cladding) & 1.200000 & 0.105503 & 1.300000 & 0.007039 & 0.023827 & 113.881899 \\
\hline
\end{tabular}
\end{table}

\subsection{Radiation patterns of a dipole at the interface on a linear power scale}
\label{sec:S1}

\begin{figure*}[ht]
 \centering
 \includegraphics[width=0.85\textwidth]{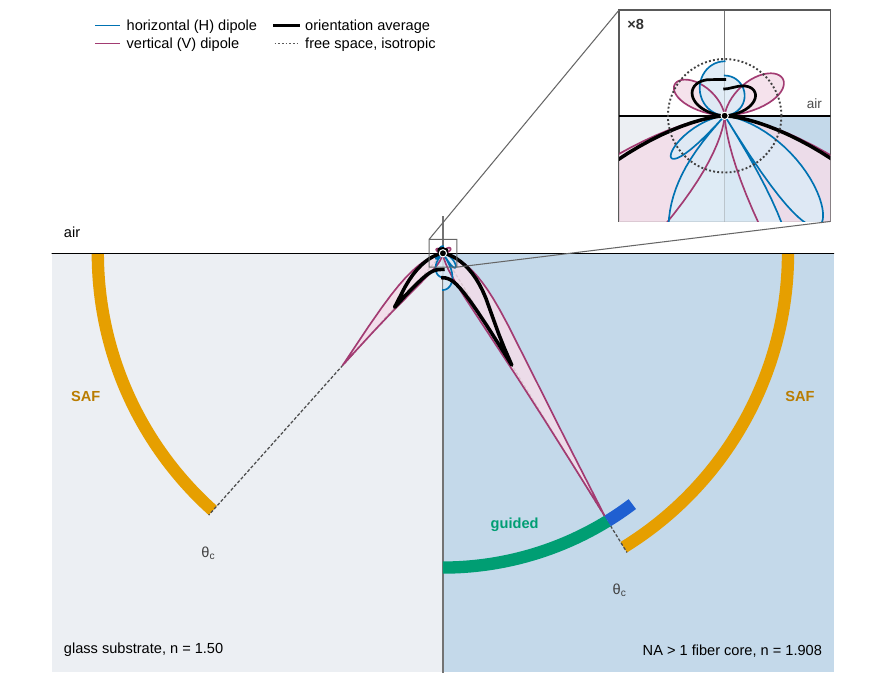}
 \caption{\textbf{Radiation patterns of Fig.~\ref{fig:1}a on a linear power scale.}
 Angular emission of an electric dipole located at the interface ($d \to 0$) between air and an ordinary glass substrate (left, $n = 1.50$) or the core of an NA\,$>$\,1 fiber (right, $n = 1.908$), calculated with the Lukosz--Kunz model~\cite{Lukosz1977_I,Lukosz1977_II,Lukosz1979_III}. Same data as in Fig.~\ref{fig:1}a, but the radial coordinate is proportional to the angular power density $\mathrm{d}P/\mathrm{d}\Omega$ instead of its square root; all curves share a common scale. Thin lines show horizontal (H) and vertical (V) dipoles; for H, the cross-section in the plane containing the dipole is shown. The thick black line is the average over random dipole orientations, $(P_\mathrm{V} + 2\langle P_\mathrm{H}\rangle_\varphi)/3$. Outer rims mark the supercritical-angle fluorescence (SAF) range, $\theta > \theta_\mathrm{c}$, and the acceptance cone of a fiber with $\mathrm{NA} = 1.15$. Inset: the region around the dipole magnified $\times 8$, showing emission into air and an isotropic emitter in free space (dotted circle, $\mathrm{d}P/\mathrm{d}\Omega = P_0/4\pi$) on the same scale.}
 \label{fig:S1}
\end{figure*}

Fig.~\ref{fig:1}a of the main text uses a radial coordinate proportional to $(\mathrm{d}P/\mathrm{d}\Omega)^{1/2}$ to make weak lobes visible. Fig.~\ref{fig:S1} shows the same data with the radial coordinate proportional to $\mathrm{d}P/\mathrm{d}\Omega$. Neither representation shows the emitted power as an area, because the power in an angular range also contains the solid-angle factor $2\pi\sin\theta$; the fractions of power in the subcritical, guided-supercritical and unguided ranges are given quantitatively by the momentum spectrum in Fig.~\ref{fig:1}b.

The critical angle $\theta_\mathrm{c} = \arcsin(1/n)$ equals \qty{41.8}{\degree} for glass and \qty{31.6}{\degree} for the fiber core, and emission at $\theta > \theta_\mathrm{c}$ (measured from the surface normal in the dielectric) constitutes SAF. The acceptance cone of the fiber, $\theta_\mathrm{a} = \arcsin(\mathrm{NA}/n_\mathrm{core})$, equals \qty{37.0}{\degree} for $\mathrm{NA} = 1.15$. Since $\theta_\mathrm{a} > \theta_\mathrm{c}$, the fiber collects part of the SAF range in addition to all subcritical emission, which a fiber with $\mathrm{NA} < 1$ cannot do by construction.

Most of the power of a near-interface dipole is radiated into the dielectric rather than into air, and a large part of it lies close to or beyond $\theta_\mathrm{c}$. The V dipole emits predominantly into the SAF range and vanishes along the normal, whereas the H dipole contributes more at subcritical angles. Increasing $n$ from 1.50 to 1.908 lowers $\theta_\mathrm{c}$ and increases the emission into the dielectric; the additional power appears almost entirely at supercritical angles (Fig.~\ref{fig:1}b). These patterns describe the limit $d \to 0$; the finite emitter--facet distance, the ND and the finite core are included in the FDTD simulations (Section~\ref{sec:S_fdtd}), which reproduce the analytical capture fractions (Fig.~\ref{fig:3}).

\subsection{FDTD simulations: overview and convergence}
\label{sec:S_fdtd}

\begin{figure*}[ht]
 \centering
 \includegraphics[width=0.85\textwidth]{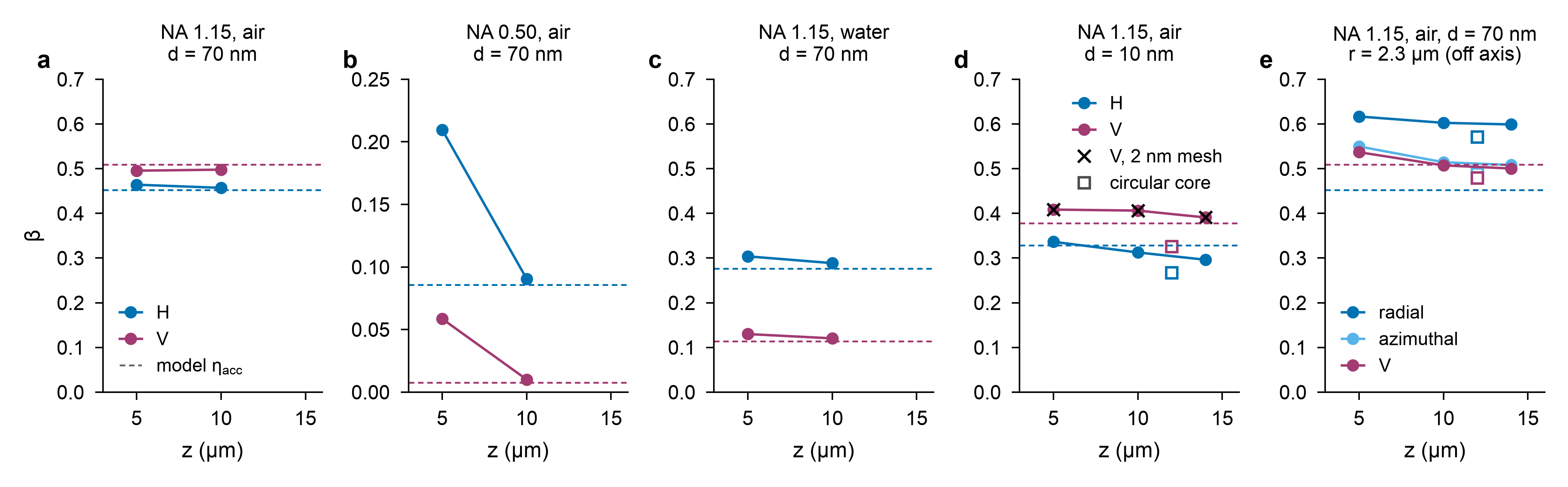}
 \caption{\textbf{Convergence of the FDTD coupling efficiency.} $\beta$ from the flux through a square enclosing the core at $z = 5$, 10 and, where available, \qty{14}{\micro\meter} behind the facet (filled circles) and through the circular core at $z = \qty{12}{\micro\meter}$ (open squares, where available), for the configurations of Table~\ref{tab:fdtd}. Dashed lines: analytical $\eta_\mathrm{acc}$ of a bare dipole at the same distance for H and V orientation. Crosses in \textbf{d}: vertical dipole with a \qty{2}{\nano\meter} instead of \qty{3}{\nano\meter} local mesh. In \textbf{e} the horizontal dipole is resolved into radial and azimuthal orientations.}
 \label{fig:S_convergence}
\end{figure*}

The simulation setup is described in Methods. Table~\ref{tab:fdtd} lists all simulations and their results. The coupling efficiency was evaluated at two or three depths behind the facet (Fig.~\ref{fig:S_convergence}). For on-axis emitters at $d = \qty{70}{\nano\meter}$ in air, $\beta$ changes by less than 1.5\% between $z = 5$ and \qty{10}{\micro\meter}. For the NA\,=\,0.5 fiber, $\beta$ is converged only at $z = \qty{10}{\micro\meter}$, because unguided rays reach the core boundary only after $\approx$\qty{7.6}{\micro\meter}. For $d = \qty{10}{\nano\meter}$, about 60\% of the power entering the glass lies beyond the fiber cutoff, and part of it travels almost parallel to the axis in the cladding; the flux through a square enclosing the core then overestimates and the flux through the circular core underestimates the guided power, and we report the range between the two. Refining the local mesh from 3 to \qty{2}{\nano\meter} for $d = \qty{10}{\nano\meter}$ changed $P_\mathrm{tot}$ by 1.3\% in both the simulation with and without the fiber, leaving $F$ and $\beta$ unchanged to within 0.1\%. All simulations reached the field-decay criterion of $10^{-6}$.

\begin{table}[ht]
\centering
\caption{FDTD simulations (ND, $\lambda = \qty{700}{\nano\meter}$, on-axis emitters unless stated otherwise; the distance given is the emitter--facet distance $d$). $\beta$ at $z = \qty{10}{\micro\meter}$; for $d = \qty{10}{\nano\meter}$ and $r = \qty{2.3}{\micro\meter}$, range between the flux through the circular core ($z = \qty{12}{\micro\meter}$) and through a square enclosing it ($z = \qty{14}{\micro\meter}$). $F$ relative to the same ND without the fiber. Model: bare dipole at the same distance.}
\label{tab:fdtd}
\small\setlength{\tabcolsep}{2pt}
\begin{tabular}{lcccccc}
\hline
Configuration & \multicolumn{3}{c}{$\beta$ (FDTD)} & $\eta_\mathrm{acc}$ (model) & \multicolumn{2}{c}{$F$ (FDTD)} \\
 & H & V & avg & avg & V & avg \\
\hline
NA 1.15, air, $d = \qty{70}{\nano\meter}$ & 0.457 & 0.498 & 0.478 & 0.478 & 2.48 & 1.61 \\
NA 0.50, air, $d = \qty{70}{\nano\meter}$ & 0.091 & 0.010 & 0.053 & 0.052 & 2.03 & 1.44 \\
NA 1.15, water, $d = \qty{70}{\nano\meter}$ & 0.289 & 0.120 & 0.220 & 0.212 & 1.45 & 1.19 \\
NA 1.15, air, $d = \qty{10}{\nano\meter}$ & 0.27--0.30 & 0.33--0.39 & 0.30--0.36 & 0.352 & 8.69 & 4.22 \\
NA 1.15, air, \qty{70}{\nano\meter}, $r = \qty{2.3}{\micro\meter}$ & 0.49--0.60$^{a}$ & 0.48--0.50 & 0.50--0.53 & 0.478$^{b}$ & 2.47 & 1.61 \\
Planar ($n = 1.908$), $d = \qty{70}{\nano\meter}$ & -- & -- & -- & -- & 2.48 & -- \\
\hline
\multicolumn{7}{l}{\footnotesize $^{a}$ azimuthal 0.49--0.51, radial 0.57--0.60. $^{b}$ bound rays only (Section~\ref{sec:S_offaxis}).}
\end{tabular}
\end{table}

\subsection{Orientation-resolved near fields}
\label{sec:S_maps}

\begin{figure*}[ht]
 \centering
 \includegraphics[width=\textwidth]{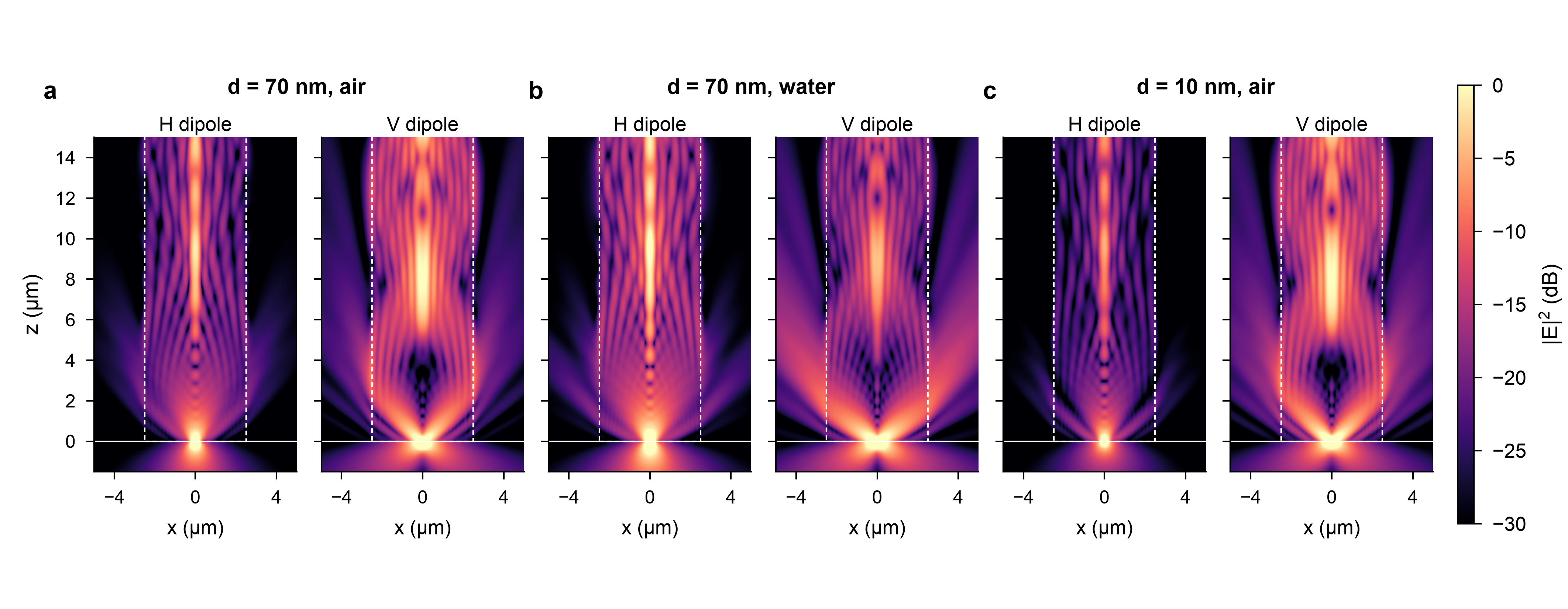}
 \caption{\textbf{Orientation-resolved field maps.} Electric-field intensity (logarithmic scale, common within each pair) in the $x$--$z$ plane for horizontal ($x$) and vertical ($z$) dipoles in a \qty{140}{\nano\meter} ND on the NA\,=\,1.15 fiber: \textbf{a},~$d = \qty{70}{\nano\meter}$ in air; \textbf{b},~$d = \qty{70}{\nano\meter}$ in water; \textbf{c},~$d = \qty{10}{\nano\meter}$ in air. Dashed lines mark the core.}
 \label{fig:S_maps}
\end{figure*}

Fig.~\ref{fig:S_maps} shows the electric-field intensity for horizontal and vertical dipoles in the ND at $d = 70$ and \qty{10}{\nano\meter}, in air and in water. For on-axis dipoles the $y$-oriented dipole is equivalent to the $x$-oriented one rotated by \qty{90}{\degree} and is not shown. Horizontal dipoles emit mainly into subcritical directions and couple to low-order guided modes, whereas vertical dipoles emit a hollow beam at supercritical angles, which the NA\,=\,1.15 fiber guides in air but not in water.

\subsection{Shallow emitters and emission enhancement at the ND--glass contact}
\label{sec:S_shallow}

For an NV center \qty{10}{\nano\meter} from the facet, i.e.\ close to the contact point between the ND and the core, the near-field interaction is strongest. The capture fraction decreases to $\beta = 0.30$--0.36, close to the analytical value of 0.35 for a bare dipole at the same distance, because a larger part of the near field is converted into wavevectors beyond the fiber cutoff. At the same time the emission rate of a vertical dipole increases by $F_\mathrm{V} = 8.7$ relative to the same ND without the fiber, three times more than predicted for a bare dipole ($F_\mathrm{V} = 2.9$). This broadband enhancement originates from the high-index ND in contact with the high-index core and does not involve any resonance. It implies a correspondingly shortened fluorescence lifetime of shallow NV centers on the fiber facet, which could be verified by time-resolved measurements. Averaged over a uniform distribution of NV centers within a \qty{140}{\nano\meter} ND, the analytical model gives an ensemble capture fraction of 0.46, close to the value for the central position.

\subsection{Emitters away from the fiber axis}
\label{sec:S_offaxis}

\begin{figure*}[ht]
 \centering
 \includegraphics[width=\textwidth]{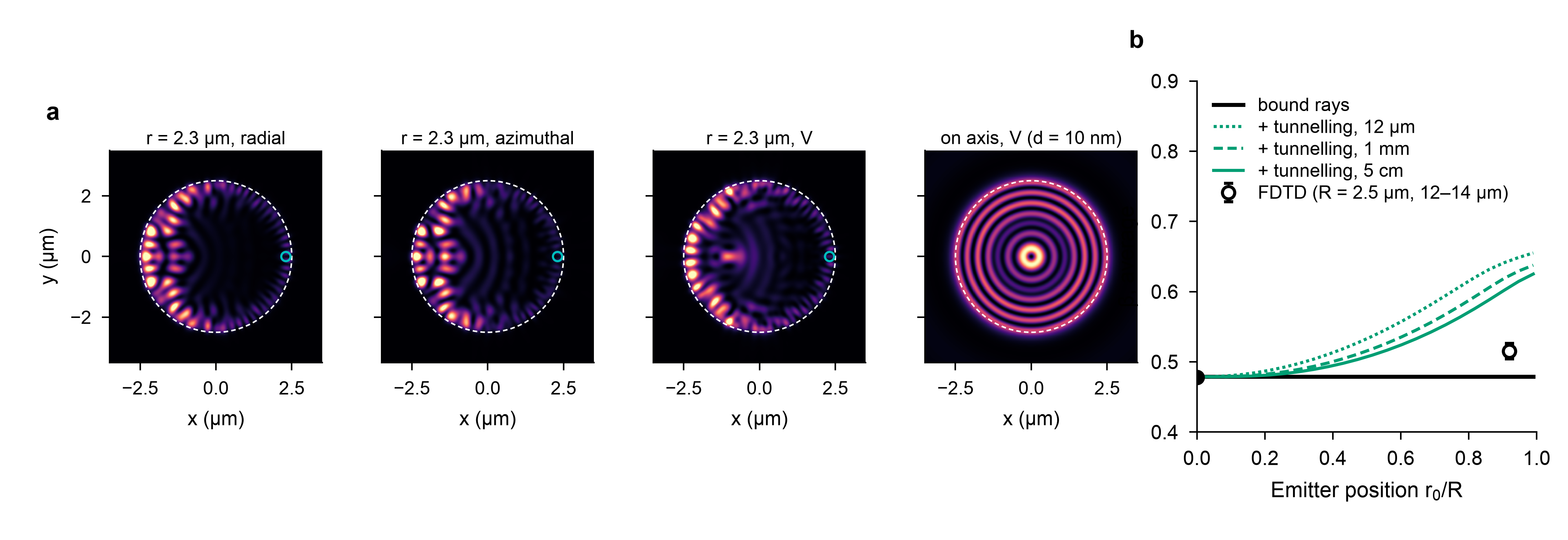}
 \caption{\textbf{Emitters away from the fiber axis.} \textbf{a},~Axial Poynting flux \qty{12}{\micro\meter} behind the facet for an ND at $r = \qty{2.3}{\micro\meter}$ (radial, azimuthal and vertical dipoles, $d = \qty{70}{\nano\meter}$; circle marks the emitter) and, for comparison, for an on-axis vertical dipole ($d = \qty{10}{\nano\meter}$); the dashed circle marks the core. \textbf{b},~Ray-optics estimate of the coupling efficiency for the \qty{10}{\micro\meter} cores of the imaging bundle as a function of the emitter position $r_0/R$: bound rays only, and bound plus tunneling rays after \qty{12}{\micro\meter}, \qty{1}{\milli\meter} and \qty{5}{\centi\meter}. Symbols: FDTD ($R = \qty{2.5}{\micro\meter}$), on the axis (filled) and at $r_0/R = 0.92$ (open; range between the flux through the circular core at \qty{12}{\micro\meter} and through the enclosing square at \qty{14}{\micro\meter}).}
 \label{fig:S_offaxis}
\end{figure*}

NDs cover the whole facet, not only the core centers. An FDTD simulation of an ND \qty{0.2}{\micro\meter} from the core edge ($r = \qty{2.3}{\micro\meter}$, core radius \qty{2.5}{\micro\meter}, $d = \qty{70}{\nano\meter}$) gives the same emission rate as on the axis ($F = 1.61$) but a higher guided fraction after \qtyrange{12}{14}{\micro\meter} of propagation ($\beta = 0.50$--0.53 instead of 0.48), with the largest increase for a radially oriented dipole. An off-axis emitter launches skew rays, part of which have $k_\parallel > \mathrm{NA}\,k_0$ but are still totally reflected at the core boundary because of their angular momentum; these tunneling leaky rays lose power slowly~\cite{SnyderLove1983} and concentrate near the core boundary (Fig.~\ref{fig:S_offaxis}a).

For a ray launched at radius $r_0$ with in-plane wavevector $q k_0$ at azimuth $\varphi$ relative to the radial direction, the ray is bound for $q < \mathrm{NA}$ and tunneling for $q^2\cos^2\varphi + q^2\sin^2\varphi\,(1 - r_0^2/R^2) < \mathrm{NA}^2 \le q^2$. Combining this criterion with the analytical angular spectrum and a Wentzel--Kramers--Brillouin (WKB) estimate of the tunneling loss per reflection gives, for the \qty{10}{\micro\meter} cores of the imaging bundle, an area-averaged coupling efficiency of at most 0.55 after \qty{5}{\centi\meter}, compared with 0.48 for bound rays alone (Fig.~\ref{fig:S_offaxis}b). The FDTD result for $r_0 = \qty{2.3}{\micro\meter}$ lies between the bound-ray value and this estimate, which therefore overstates the tunneling contribution near the core edge; $\beta \approx 0.5$ is a realistic value for emitters distributed over the core.

Emitters above the cladding cannot excite bound or tunneling core modes in the ray picture, because rays entering the cladding have axial wavevectors below $n_\mathrm{clad}k_0$; they couple only through the evanescent tails of the guided modes within about one wavelength of the core edge. With a core fill factor of about 23\% (core diameter \qty{10}{\micro\meter}, pitch \qty{20}{\micro\meter}), about 11--12\% of the emission of all NDs on the facet is coupled into core modes, while emitters above the cladding mainly add background in cladding modes. Tunnelling rays also leave a short fiber at angles beyond the nominal NA, which may contribute to the slightly larger NA measured in Section~\ref{sec:S_NA}.

\subsection{Aqueous environment}
\label{sec:S_water}

\begin{figure*}[ht]
 \centering
 \includegraphics[width=0.6\textwidth]{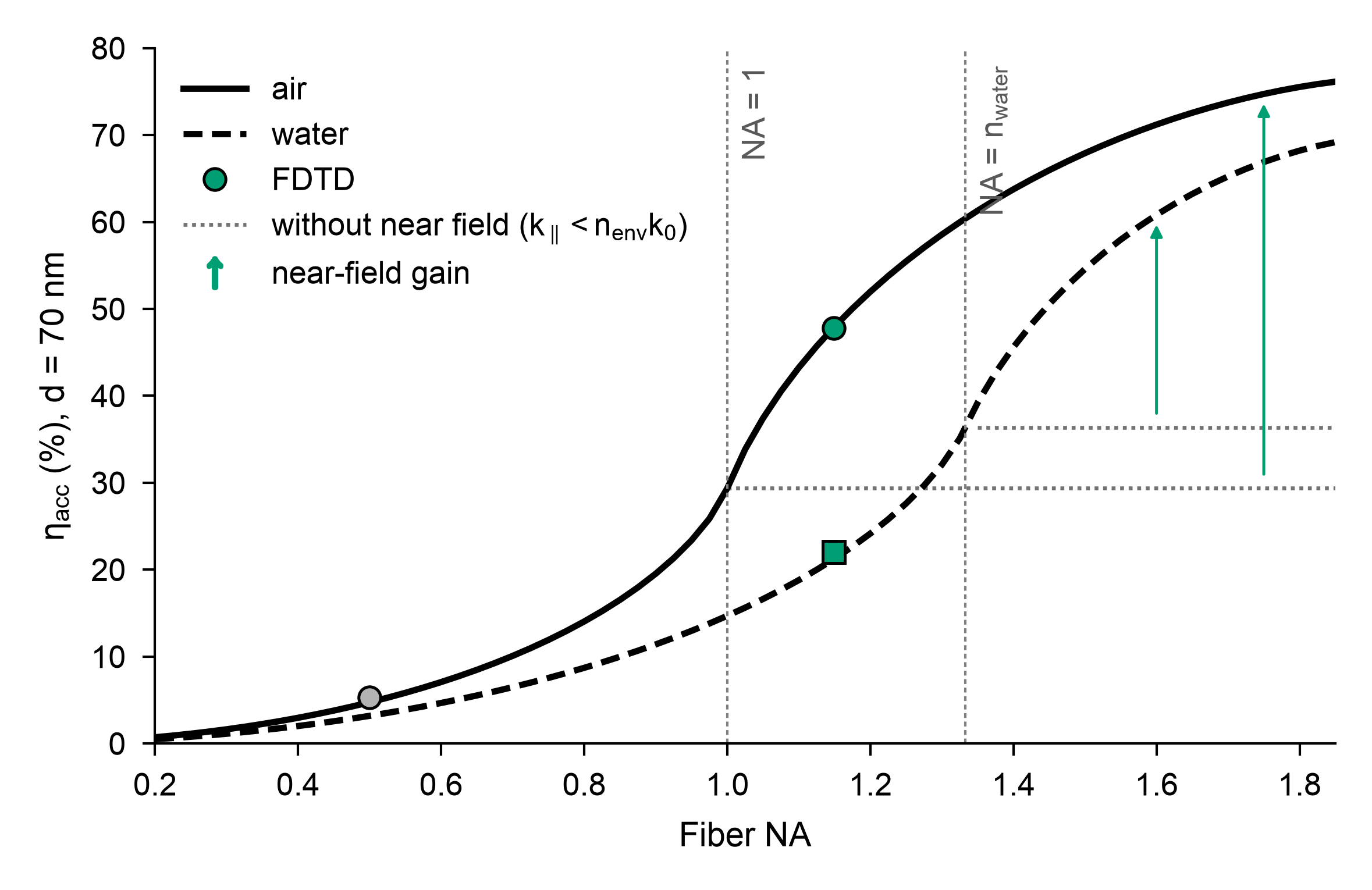}
 \caption{\textbf{Capture in air and in water.} Orientation-averaged $\eta_\mathrm{acc}$ ($d = \qty{70}{\nano\meter}$, $n_\mathrm{core} = 1.908$) as a function of NA in air (solid) and in water (dashed). Dotted lines: capture without the near-field contribution ($k_\parallel < n_\mathrm{env}k_0$), which saturates at $\mathrm{NA} = n_\mathrm{env}$; arrows: near-field gain. Symbols: FDTD with the ND (circle, air; square, water; grey circle, NA\,=\,0.5 fiber with $n_\mathrm{core} = 1.60$).}
 \label{fig:S_water}
\end{figure*}

In water, the fiber NA of 1.15 is smaller than the refractive index of the environment ($n_\mathrm{env} = 1.33$). The acceptance window $k_\parallel < \mathrm{NA}\,k_0$ then lies entirely below the light line of water, all guided light originates from waves that propagate in water, and the guided emission rate becomes independent of the emitter distance. The orientation-averaged capture fraction decreases to $\eta_\mathrm{acc} = 0.21$ (FDTD: $\beta = 0.22$), mainly because vertical dipoles are collected poorly ($\beta_\mathrm{V} = 0.12$), and the emission rate is only weakly modified ($F = 1.19$). Near-field capture in water requires $\mathrm{NA} > 1.33$: for $d = \qty{70}{\nano\meter}$ the analytical model gives $\eta_\mathrm{acc} = 0.36$ at $\mathrm{NA} = 1.33$ and 0.55 at $\mathrm{NA} = 1.5$, compared with 0.60 and 0.68 in air (Fig.~\ref{fig:S_water}). For the present core glass ($n_\mathrm{core} = 1.908$), a cladding index of 1.40, 1.36 or 1.31 gives $\mathrm{NA} = 1.30$, 1.34 or 1.39 and $\eta_\mathrm{acc} = 0.32$, 0.37 or 0.44 in water; a suspended core gives $\mathrm{NA} = 1.37$ ($\eta_\mathrm{acc} = 0.41$) if its air holes fill with water at the facet and $\mathrm{NA} = 1.62$ ($\eta_\mathrm{acc} = 0.62$) if they are sealed. The emission rate is the same for all these designs, since it depends only on $n_\mathrm{core}/n_\mathrm{env}$. In air, the cladding index of 1.31 gives $\eta_\mathrm{acc} = 0.63$. An air-clad polymer fiber with a poly(methyl methacrylate) core ($n = 1.49$) would, if its NA reached 1, collect the entire subcritical cone, $\eta_\mathrm{acc} = 0.34$ in air at $d = \qty{70}{\nano\meter}$, six times more than an NA\,=\,0.5 fiber.

\subsection{Fiber structure and experimental setup}
\label{sec:S_setup}

\begin{figure*}[t]
    \centering
    \includegraphics[width=1\textwidth]{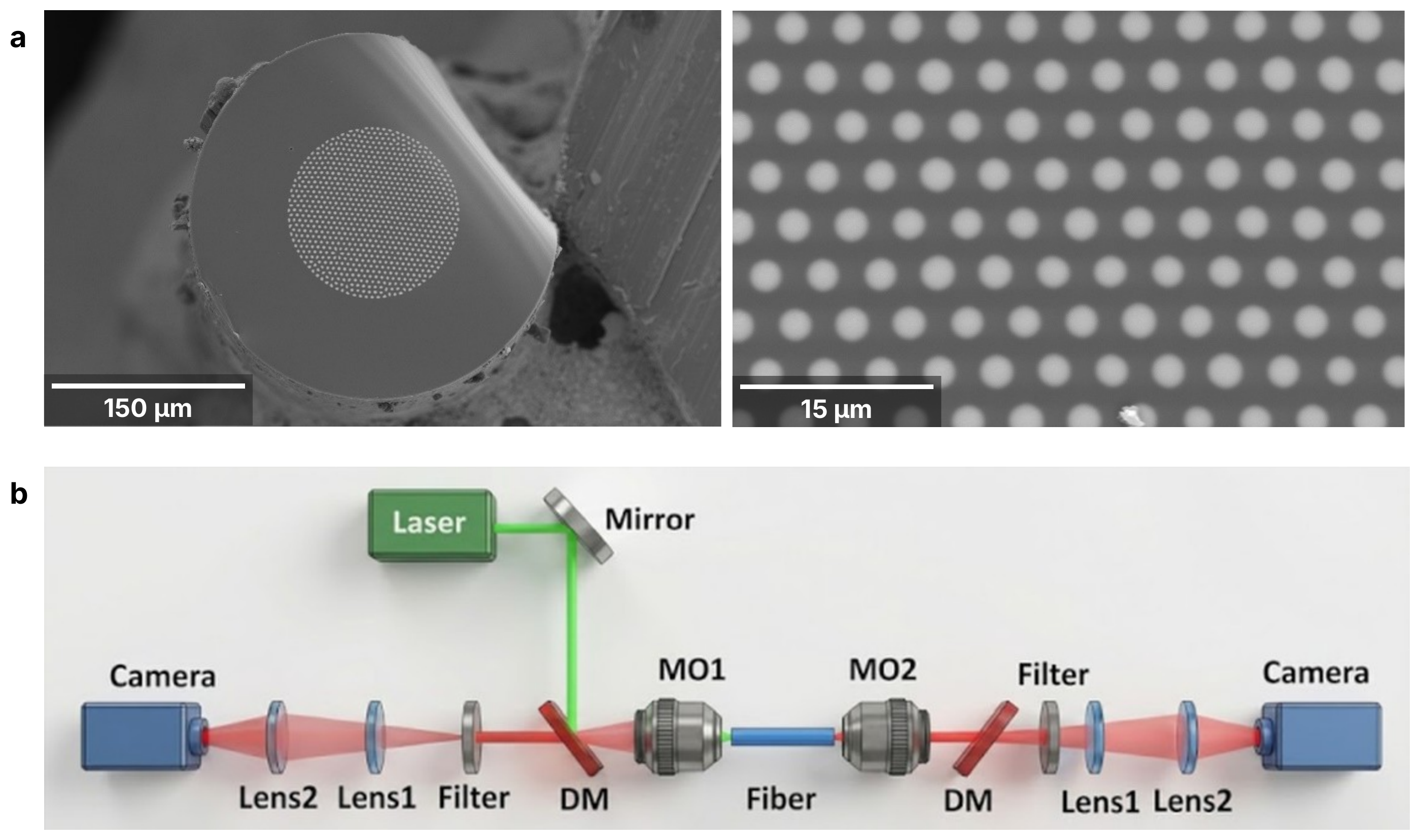}
    \caption{\textbf{a}, SEM images of imaging fiber bundle HC1\_2\_6 (Table~\ref{tab:bundles}), drawn from the same glasses as the endoscope bundle: facet (left) and close-up of the core area (right). \textbf{b}, Schematic of the back-focal-plane imaging setup. MO1, excitation objective facing the coated facet; MO2, readout objective with variable aperture; DM, dichroic mirror.}
    \label{fig:Fiber-setup}
\end{figure*}

The endoscope bundle, first used in Ref.~\cite{Wojciechowski2025SPIE}, belongs to a series of rigid bundles drawn from the same pair of glasses: the lead--bismuth--gallate glass PBG81 for the cores and the borosilicate glass UV710 for the cladding, arranged as in Ref.~\cite{Bijoch2023} (Fig.~\ref{fig:Fiber-setup}a). Its cores are about \qty{10}{\micro\meter} wide and spaced by \qty{19.2}{\micro\meter}, and about 830 of them fill an imaging area \qty{578}{\micro\meter} across (Methods). From the refractive indices at \qty{700}{\nano\meter}, $n_\mathrm{core} = 1.908$ and $n_\mathrm{clad} = 1.524$ (Section~\ref{sec:S_dispersion}), $\mathrm{NA} = \sqrt{n_\mathrm{core}^2 - n_\mathrm{clad}^2} = 1.15$. A \qty{5}{\centi\meter} piece of this bundle carries on one facet a layer of \qty{140}{\nano\meter} NDs (Adamas Nanotechnologies), bound covalently following Refs.~\cite{Czarnecka2022, Jani2023}.

Bundles with smaller cores have been drawn from the same core and cladding glasses (Table~\ref{tab:bundles}); the \qty{3.5}{\centi\meter}-long bundle used for back-focal-plane imaging is HC1\_2\_4.

\begin{table}[ht]
\centering
\caption{Geometry of imaging fiber bundles drawn from the same core and cladding glasses (ranges over the facet). For a core diameter $2a$, the number of guided modes at \qty{700}{\nano\meter} and $\mathrm{NA} = 1.15$ is approximately $V^2/2$ with $V = 2\pi a\,\mathrm{NA}/\lambda$.}
\label{tab:bundles}
\small
\begin{tabular}{lccccc}
\hline
Bundle & Outer & Imaging & Core & Pitch & Modes \\
 & diam.\ (\unit{\micro\meter}) & diam.\ (\unit{\micro\meter}) & diam.\ (\unit{\micro\meter}) & (\unit{\micro\meter}) & per core \\
\hline
HC1\_2\_0 & 853 & 406--408 & 5.8--6.6 & 13.4--14.0 & 510 \\
HC1\_2\_3 & 914 & 428--431 & 7.5--8.0 & 14.3--14.8 & 800 \\
HC1\_2\_4 & 567 & 267 & 4.8--5.2 & 8.7--9.1 & 330 \\
HC1\_2\_6 & 281 & 132--133 & 2.6--2.8 & 4.2--4.5 & 97 \\
HC1\_2\_IV & 265 & 123--124 & 2.3--2.6 & 3.9--4.3 & 79 \\
HC1\_2\_V & 243 & 113--114 & 2.4--2.6 & 3.7--3.8 & 85 \\
HC1\_3\_A & 137 & 65 & 1.3--1.4 & 2.1--2.2 & 25 \\
HC1\_3\_B & 164 & 76--77 & 1.9--2.1 & 2.5--2.6 & 52 \\
HC1\_3\_E & 127 & 59 & 1.4--1.5 & 1.8--2.0 & 27 \\
\hline
\end{tabular}
\end{table}

The lead--bismuth--gallate core introduces background luminescence under \qty{532}{\nano\meter} excitation. This autofluorescence, originating from trace impurities or structural defects of the heavy-metal oxide glass, spans a broad spectral range overlapping with the NV phonon sideband. As noted in previous studies on fiber-coupled quantum emitters~\cite{Schroder2012}, fiber background fluorescence can reduce the signal-to-noise ratio and the measurement contrast. Because it is generated inside the core, it fills the fiber acceptance without the angular signature of SAF (Section~\ref{sec:S_bfp}). Future high-index probes may benefit from glasses of higher purity or from time-gated detection separating the NV luminescence ($\tau \approx 10$--\qty{20}{\nano\second}) from the prompt glass autofluorescence.

\subsection{Numerical aperture measurement}
\label{sec:S_NA}

\begin{figure*}[t]
    \centering
    \includegraphics[width=0.75\textwidth]{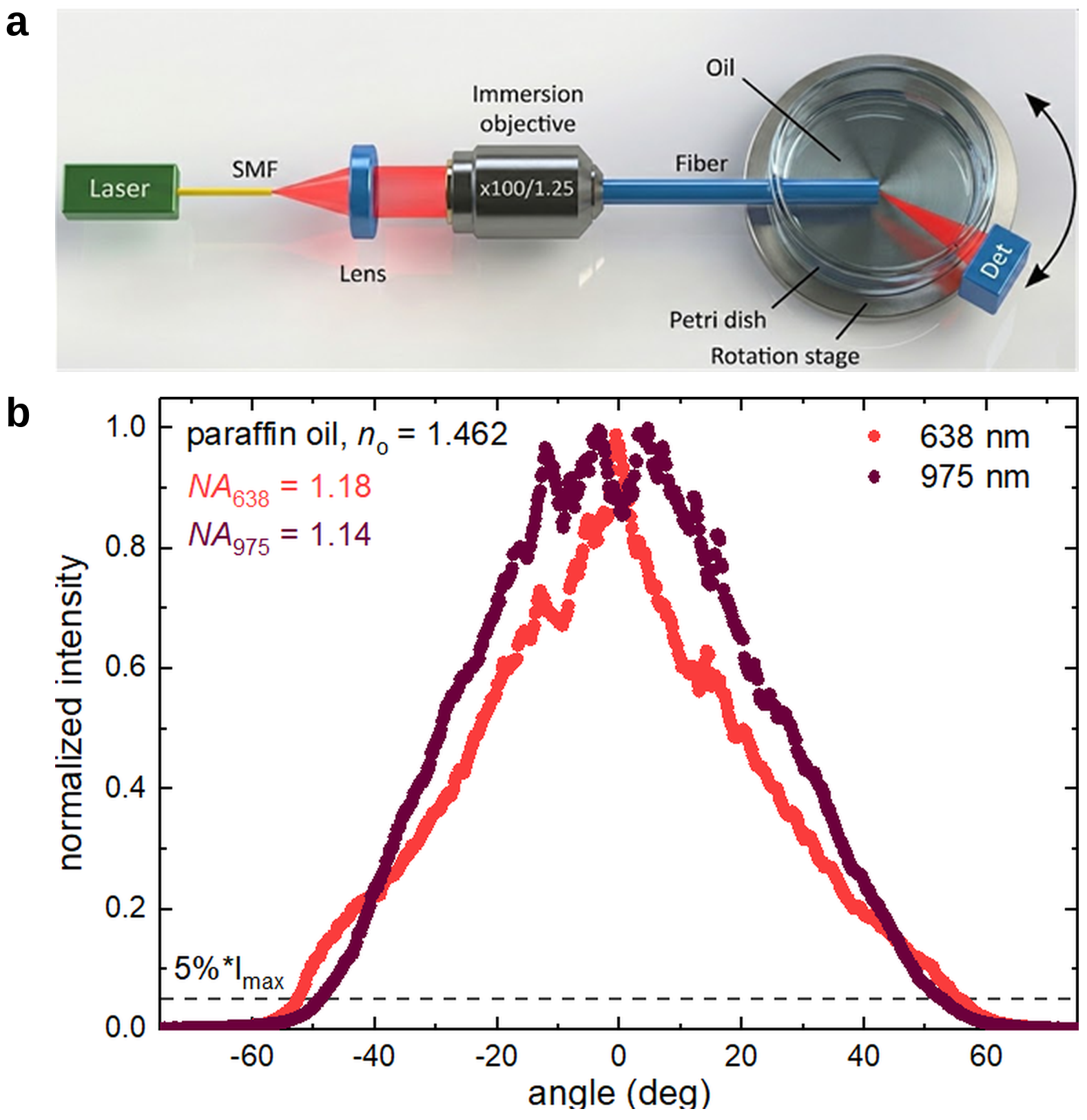}
    \caption{\textbf{a}, Schematic of the measurement system used to verify the numerical aperture (NA) of the optical fiber. \textbf{b}, Normalized angular intensity distribution measured in paraffin oil for wavelengths \qty{638}{\nano\meter} (red) and \qty{975}{\nano\meter} (purple). The resulting NA values are indicated.}
    \label{fig:NA-setup}
\end{figure*}

To verify the numerical aperture of the fiber, a custom immersion-based goniometric system was used (Fig.~\ref{fig:NA-setup}a). Two switchable fiber-pigtailed laser diodes, operating at \qty{638}{\nano\meter} and \qty{975}{\nano\meter}, served as light sources. The diode output was collimated by an aspheric lens ($f = \qty{18}{\milli\meter}$) and focused by a $100\times/1.25$ oil-immersion microscope objective immersed in paraffin oil ($n = 1.462$) to couple light into the fiber across the full numerical aperture of the objective. Both the objective and the fiber input end were submerged in the oil.

The fiber output end was positioned at the center of a cylindrical \qty{20}{\centi\meter}-diameter glass dish filled with paraffin oil, which suppresses total internal reflection at the output facet and reduces the divergence of the emerging beam. A motorized rotation stage (Thorlabs PRM1Z8), whose axis coincided with the fiber output tip, was mounted beneath the dish, and a photodiode power sensor (Thorlabs S120VC) behind a \qty{2}{\milli\meter}-wide slit was attached to an arm on the stage, \qty{1}{\centi\meter} outside the dish wall. Because the output facet was located at the center of curvature of the cylindrical wall, rays from the fiber tip crossed the oil--glass--air boundary at normal incidence and the angular scan directly measured the in-oil propagation angle $\theta$ (Fig.~\ref{fig:NA-setup}b).

The maximum in-oil emission angle was defined by the intensity falling to 5\% of its maximum. The resulting values, $\sin\theta_\mathrm{oil,max} = 0.81$ and 0.78 at \qty{638}{\nano\meter} and \qty{975}{\nano\meter}, give $\mathrm{NA} = n\sin\theta_\mathrm{oil,max} = 1.18$ and 1.14, respectively. The values calculated from the Sellmeier coefficients are 1.160 and 1.124 (Fig.~\ref{fig:S_dispersion}a); the measured NA is thus about 0.02 higher at both wavelengths, which is within the uncertainty of the threshold definition and consistent with a small contribution of tunneling leaky rays in a short fiber (Section~\ref{sec:S_offaxis}).

\subsection{Back-focal-plane imaging: model and calibration}
\label{sec:S_bfp}

\begin{figure*}[ht]
 \centering
 \includegraphics[width=0.85\textwidth]{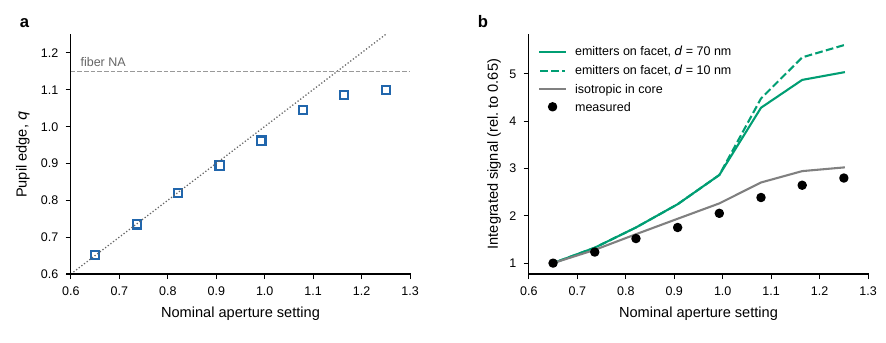}
 \caption{\textbf{Calibration of the back-focal-plane images and comparison with the model.} \textbf{a},~Pupil edge (half-maximum radius of the azimuthally averaged profile) as a function of the nominal aperture setting, converted to $q$ with the scale obtained from the three smallest settings (\qty{388}{px} per unit NA). Dotted line: edge equal to the nominal setting; dashed line: fiber NA. \textbf{b},~Integrated BFP signal relative to the smallest setting: measured (circles) and predicted by the analytical model for orientation-averaged emitters on the facet ($d = 70$ and \qty{10}{\nano\meter}) and for isotropic emission inside the core, each integrated up to the measured pupil edge in \textbf{a}.}
 \label{fig:S_bfp_cal}
\end{figure*}

For an oil-immersion objective fulfilling the Abbe sine condition, the radius in the back focal plane (BFP) is proportional to the in-plane wavevector, $\rho = f\,k_\parallel/k_0$, which is conserved at the immersed output facet. In a straight, rigid fiber without mode mixing, every guided ray keeps its $k_\parallel$, so that the BFP intensity per unit area is proportional to $(1/q)\,\mathrm{d}P/\mathrm{d}q$ of the light entering the core, truncated at the fiber NA. The aperture stop of the UPLFLN 60XOI objective lies close to its back focal plane, so that, as long as the aperture limits the pupil, the pupil radius is proportional to the aperture setting. The pupil edges (half-maximum points of the azimuthally averaged profiles) at the three smallest settings (0.65, 0.74 and 0.82) give a common scale of \qty{388}{px} per unit NA, to within 1\%. With this scale the edge follows the nominal setting up to about 1 and saturates at $q = 1.10$ for settings of 1.16 and 1.25 (Fig.~\ref{fig:S_bfp_cal}a), 4\% below the fiber NA; the residual difference is within the uncertainty of the nominal aperture scale and the width of the pupil edge (about 25 pixels, i.e.\ 0.065 in $q$). An earlier series, recorded before the Bertrand lens was refocused onto the back focal plane, showed a pupil edge saturating at $q \approx 0.93$ and a bright rim inside the edge at every aperture setting, which we attribute to the misadjusted relay; it was not used.

\begin{figure*}[ht]
 \centering
 \includegraphics[width=0.85\textwidth]{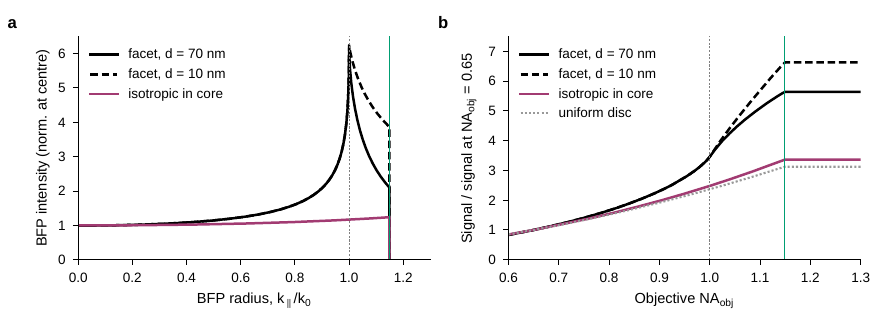}
 \caption{\textbf{Expected back-focal-plane signal.} \textbf{a},~BFP intensity per unit area as a function of $k_\parallel/k_0$ for orientation-averaged dipoles on the facet ($d = 70$ and \qty{10}{\nano\meter}, analytical model) and for isotropic emission inside the core, both truncated at $\mathrm{NA} = 1.15$ and normalized at the center. \textbf{b},~Integrated signal as a function of the objective NA, normalized at $\mathrm{NA}_\mathrm{obj} = 0.65$, compared with a uniformly filled aperture. }
 \label{fig:S_bfp}
\end{figure*}

Fig.~\ref{fig:S_bfp} compares the expected BFP profiles for emitters on the facet with those for an isotropic source inside the core, such as glass autofluorescence. Emission from the facet produces a sharp maximum at $k_\parallel/k_0 = 1$, independent of the emitter distance, orientation and wavelength, followed by a decrease towards the cutoff; isotropic emission inside the core produces a nearly uniform disc, brightening monotonically by a factor of 1.24 towards the cutoff and without any feature at the light line. The integrated signal increases from $\mathrm{NA}_\mathrm{obj} = 0.65$ to saturation by a factor of 5.6 ($d = \qty{70}{\nano\meter}$) to 6.6 ($d = \qty{10}{\nano\meter}$) for emitters on the facet, and by 3.4 for isotropic emission in the core.

Evaluated up to the measured pupil edge at each setting (saturating at $q = 1.10$ instead of 1.15), the predicted increase between the settings 0.65 and 1.25 is 5.0 ($d = \qty{70}{\nano\meter}$) to 5.6 ($d = \qty{10}{\nano\meter}$) for emitters on the facet and 3.0 for isotropic emission; the measured increase is 2.8 (Fig.~\ref{fig:S_bfp_cal}b). The measured profiles (Fig.~\ref{fig:4}d) show a maximum at $q \approx 1.04$, about 20\% above the plateau, whereas the model with the same edge width predicts a maximum about four times the central intensity for emitters on the facet. The intensity also decreases by about 30\% from the center of the pupil to $q \approx 0.85$, whereas both model profiles are flat or increase towards the edge. Both observations indicate angle-dependent losses not included in the model and a background that fills the pupil uniformly. A reference measurement on an uncoated fiber would separate these contributions.

The same model predicts the signal recorded at the output facet with different readout objectives (Fig.~\ref{fig:4}a,b), which does not require a calibration of the pupil. A dry objective of numerical aperture $\mathrm{NA}_\mathrm{obj}$ collects the guided light with $q \le \mathrm{NA}_\mathrm{obj}$, weighted by the unpolarized Fresnel transmission of the glass--air output facet, whereas an oil-immersion objective ($n = 1.518$) collects all guided light up to the fiber NA with much higher transmission. For isotropic emission inside the core, the guided power per unit $q$ is proportional to $q/(n_\mathrm{core}^2 - q^2)^{1/2}$. Table~\ref{tab:readout} compares both predictions with the immersion/dry gain measured with the zero-field resonance (Fig.~\ref{fig:4}b and Section~\ref{sec:S_airoil}). Only the dry-to-immersion ratio, which is dominated by guided light with $q > 0.95$, distinguishes clearly between the two sources.

\begin{table}[ht]
\centering
\caption{Signal ratios for different readout objectives at the output facet, predicted by the analytical model (conservation of $k_\parallel$ along the fiber, Fresnel transmission at the output facet), and the immersion/dry gain measured with the zero-field resonance (Fig.~\ref{fig:4}b): resonant PL and, in parentheses, total PL of the cores ($\pm 0.03$ and $\pm 0.05$).}
\label{tab:readout}
\begin{tabular}{lccc}
\hline
Readout objectives & Measured & Facet emitters, $d = 70$ (10)\,nm & Isotropic in core \\
\hline
NA 0.95 dry / NA 0.6 dry & -- & 3.2 (3.2) & 2.6 \\
NA 1.4 oil / NA 0.95 dry & 1.86 (1.58) & 2.3 (2.7) & 1.7 \\
NA 1.4 oil / NA 0.6 dry & -- & 7.4 (8.7) & 4.3 \\
\hline
\end{tabular}
\end{table}

\subsection{Oersted field of the test structure}
\label{sec:S_oersted}

\begin{figure*}[ht]
\centering
\includegraphics[width=\textwidth]{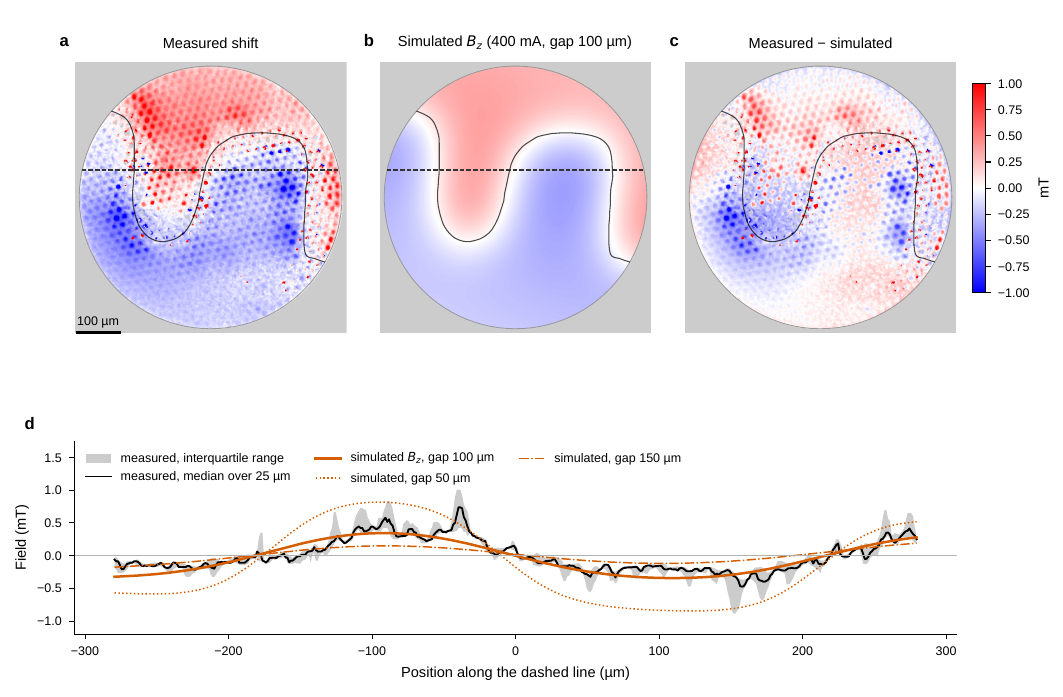}
\caption{\textbf{Comparison of the measured field map with the Oersted field of the trace.} \textbf{a},~Measured position of the resonance (Fig.~\ref{fig:endoscope}b). \textbf{b},~Calculated component $B_z$ along the fiber axis for \qty{400}{\milli\ampere} at a distance of \qty{100}{\micro\meter} above the trace, plus the fitted offset. \textbf{c},~Difference between \textbf{a} and \textbf{b}. Lines: current path used in the calculation. \textbf{d},~Profiles along the dashed line in \textbf{a} and \textbf{b}: measured (median and interquartile range over a band \qty{25}{\micro\meter} wide) and calculated for distances of 50, 100 and \qty{150}{\micro\meter}. Scale bar, \qty{100}{\micro\meter}.}
\label{fig:S_oersted}
\end{figure*}

The current path was obtained from the confocal image of the test structure as the shortest path along the skeleton of the segmented trace between the two leads, which excludes the lower arc of the `J' that carries no current. It was mapped onto the coordinates of the field maps with the same transformation as the outline in Fig.~\ref{fig:endoscope} and extended by straight leads of \qty{20}{\milli\meter}. The trace was modelled as a ribbon \qty{50}{\micro\meter} wide and \qty{35}{\micro\meter} thick, represented by $5 \times 3$ filaments carrying a total current of \qty{400}{\milli\ampere}, and the field was calculated with the Biot--Savart law in a plane at a distance $h$ above the top of the trace. Because the bias field was swept along the fiber axis, the position of the resonance measures the component $B_z$ along this axis; the in-plane components do not shift the resonance but reduce its depth, because the sweep along the axis cannot cancel them. With the amplitude fixed by the current and only a constant offset fitted, the agreement with the measured map is best for $h \approx \qty{100}{\micro\meter}$ ($R^2 = 0.51$ for the map median-filtered over 5 pixels; Fig.~\ref{fig:S_oersted}); a free amplitude equals the calculated one for $h \approx \qty{95}{\micro\meter}$. The fitted offset, \qty{+0.028}{\milli\tesla}, is of the order of the residual laboratory field. The residuals are dominated by individual cores above the conductor, where the resonance is distorted, and by a smooth variation of a few tenths of a millitesla across the bundle, which may arise from the tilt of the bundle with respect to the structure and from currents in the leads outside the field of view.

\subsection{Sensitivity estimation}
\label{sec:S_sensitivity}

The sensitivity was obtained from the slope of the resonance and the measured noise level, that is, for a measurement at the steepest point of the resonance:
\begin{equation}
\eta_B = \frac{\delta S \sqrt{t}}{S_\mathrm{max}}, \qquad S_\mathrm{max} = \sqrt{8\ln 2}\, e^{-1/2}\, \frac{A}{w} \approx 1.43\,\frac{A}{w},
\label{eq:eta_slope}
\end{equation}
where $S$ is the normalized signal (reference/signal), $A$ and $w$ are the height and full width at half maximum of the fitted Gaussian peak of $S$, $S_\mathrm{max}$ is its largest slope, $\delta S$ is the standard deviation of $S$ at one field step, and $t$ is the time spent on one field step.

For every binned pixel and field step, the standard error of $S$ was obtained from the scatter between the 16 repetitions of the sweep. Away from the trace, its median, $1.3 \times 10^{-3}$ at a PL level of about \qty{1700}{\ADU} per camera pixel, corresponds to photon shot noise for a conversion factor $K = \qty{3.0}{\electron\per\ADU}$, close to the value of \qty{2.6}{\electron\per\ADU} obtained from the saturation capacity of the sensor (\qty{10.5}{ke^-}, manufacturer's data) and its 12-bit range. The measurement is therefore shot-noise-limited at the level of a binned pixel.

Table~\ref{tab:sensitivity} lists median values for 126--193 square regions of different size located at least \qty{37}{\micro\meter} from the conductor, where the field gradient is small. The largest region, $14 \times 14$ binned pixels (\qty{302}{\micro\meter\squared}), corresponds approximately to one cell of the core lattice ($(\sqrt{3}/2)\,p^2 = \qty{319}{\micro\meter\squared}$ for $p = \qty{19.2}{\micro\meter}$). For each region the spectra were averaged and fitted by weighted least squares, and $\delta S$ was taken from the scatter of the second differences of the averaged spectrum, which is insensitive to the smooth line shape and includes any noise common to the pixels of the region. It agrees to within 20\% with the shot noise of the ratio, $\sqrt{2/N}$ for $N$ photoelectrons in each of the signal and reference images. The time per field step is either the exposure time of the signal and reference images, $t = 16 \times 2 \times \qty{70}{\milli\second} = \qty{2.24}{\second}$, or the total measurement time divided by the 80 steps, $t = \qty{31.9}{\second}$. With shot noise alone, $\delta S\sqrt{t} = 2/\sqrt{I_0}$ for the exposure time, where $I_0$ is the detected photoelectron rate, and Eq.~\eqref{eq:eta_slope} becomes $\eta_B \approx 1.4\,w/(C\sqrt{I_0})$. Because the noise is close to shot noise, $\eta_B\sqrt{A}$ hardly depends on the size of the region.

The field maps of Fig.~\ref{fig:endoscope} were obtained by fitting the whole sweep. Their statistical uncertainty $\sigma_B$ (weighted fits, scaled by the square root of the reduced $\chi^2$ where it exceeds unity) is listed for comparison; $\sigma_B\sqrt{T}$, with $T$ the total measurement time, is 2 to 3.4 times larger than $\eta_B$. A sweep spends most of its time at fields far from the flanks of the resonance, which carry little information on its position, and the reduced $\chi^2$, which increases from 1.1 for single binned pixels to 3.2 for a core cell, shows deviations from a single Gaussian line that affect fits of the whole sweep but not a measurement on the flank. Detection at the flanks, for example with a modulated bias field, and a higher duty cycle of the camera (7\% here) would realize the sensitivity of Table~\ref{tab:sensitivity} in the maps.

\textbf{Parallel readout.} The values in Table~\ref{tab:sensitivity} refer to a single region, but they hold for every region of the facet at the same time, because each camera frame records all cores of the bundle. For the largest region, one cell of the core lattice, the bundle thus provides about 800 independent field measurements, each with $\eta_B \approx \qty{0.23}{\milli\tesla\per\sqrt{\hertz}}$ over the measurement time. A single-point fiber probe with the same sensitivity would need about 800 times longer to record the same map by scanning, or, in the same total time, would reach a sensitivity per point about $\sqrt{800} \approx 28$ times worse. The modest sensitivity per cell is therefore partly offset by the number of cells read out in parallel, and the area-normalized sensitivity $\eta_B\sqrt{A}$, not $\eta_B$ alone, is the figure of merit for comparison with other wide-field imagers.

\textbf{Camera limits.} The camera (Basler acA2440-35um; \qty{3.45}{\micro\meter} pixels, saturation capacity \qty{10.5}{ke^-}, at most 35 frames per second at full resolution) can record at most $10.5\,\mathrm{ke^-} \times 35\,\unit{\per\second} = 3.7 \times 10^{5}$ photoelectrons per second per pixel, and $1.8 \times 10^{12}$\,\unit{\per\second} over the whole sensor \cite{Wojciechowski2018}. Table~\ref{tab:camera} compares this with the photoelectron rates detected during the exposures. With the $20\times$ objective, the bundle delivered $1.6 \times 10^{11}$ photoelectrons per second, below the capacity of the camera; the measurement was limited by the duty cycle of the acquisition (7\%), which left the camera idle while the coil currents were switched and settled. With the $40\times$ objectives the bundle covered fewer pixels (1.8--1.9\,Mpx instead of 2.6\,Mpx) and delivered more light: with immersion readout, $1.1 \times 10^{12}$ photoelectrons per second at 30\% of the maximum LED current, which exceeds the $6.5 \times 10^{11}$\,\unit{\per\second} that the pixels covering the bundle can record; at full LED current it would exceed the capacity of the whole sensor. In this configuration the camera, and not the fiber, sets the photon budget. Averaged over the measurement time, all three measurements recorded only about 1\% of the capacity of the pixels covering the bundle. Binning does not cost resolution here: one core pitch spans 13--15 binned pixels (52--62 camera pixels), so about $10^{4}$ pixels would suffice to sample the 830 cores, whereas the sensor has $5 \times 10^{6}$ pixels of small capacity. A scientific CMOS camera with \qty{30}{ke^-} pixels at 100 frames per second can record $1.3 \times 10^{13}$ photoelectrons per second over its sensor, enough for the full LED signal, and cameras designed for a high full-well capacity (\qty{2}{Me^-} per pixel at 550 frames per second, $2.3 \times 10^{15}$\,\unit{\per\second}) would accept laser excitation two to three orders of magnitude stronger.

\begin{table}[ht]
\centering
\caption{Photoelectron rates in the magnetic measurements and the capacity of the camera (\qty{10.5}{ke^-} $\times$ 35 frames per second per pixel). Rates during the exposures, from the mean photoluminescence and $K = \qty{2.56}{\electron\per\ADU}$; ``full LED'' extrapolates linearly to the maximum LED current of \qty{1}{\ampere}.}
\label{tab:camera}
\small
\begin{tabular}{lccc}
\hline
 & $20\times$, dry & $40\times$, dry & $40\times$, oil \\
\hline
LED current (A) & 0.9 & 0.3 & 0.3 \\
Exposure per frame (ms) & 70 & 10 & 5 \\
Pixels on the bundle (Mpx) & 2.6 & 1.9 & 1.8 \\
Brightest pixels / full scale & 0.74 & 0.62 & 0.62 \\
Rate from the bundle (\unit{\per\second}) & $1.6 \times 10^{11}$ & $7.6 \times 10^{11}$ & $1.1 \times 10^{12}$ \\
\quad at full LED current & $1.8 \times 10^{11}$ & $2.5 \times 10^{12}$ & $3.6 \times 10^{12}$ \\
Capacity of these pixels (\unit{\per\second}) & $9.5 \times 10^{11}$ & $7.1 \times 10^{11}$ & $6.5 \times 10^{11}$ \\
Duty cycle & 7.0\% & 1.3\% & 0.7\% \\
Recorded fraction of capacity$^{a}$ & 1.2\% & 1.4\% & 1.1\% \\
\hline
\end{tabular}
\par\smallskip\footnotesize $^{a}$ Averaged over the whole measurement time.
\end{table}

\textbf{Comparison with direct imaging.} Two measurements of Ref.~\cite{Sengottuvel2025}, in which the same \qty{140}{\nano\meter} NDs were drop-cast on a polymer film above a current-carrying pattern and imaged directly through a dry $40\times$/0.65 objective with the same camera and LED, were processed with the same pipeline ($4 \times 4$ binning, reference/signal ratio, weighted Gaussian fits of spectra averaged over $3 \times 3$ binned pixels; \qty{0.62}{\micro\meter} per binned pixel, exposure \qty{20}{\milli\second}, 3 and 10 sweeps). Table~\ref{tab:freespace} compares the medians in the quieter half of the field maps with the measurements through the bundle, all for spectra averaged over $3 \times 3$ binned pixels. The sensitivity was obtained with Eq.~\eqref{eq:eta_slope} from the averaged spectra of square regions in the same way as for Table~\ref{tab:sensitivity}. In all measurements the noise of $3 \times 3$ regions agrees with the photon shot noise to within 10\%. The area-normalized sensitivity is given for the exposure time, which removes differences in the duty cycle of the acquisitions; it hardly depends on the region size (Table~\ref{tab:sensitivity}). Both the bundle and the direct imaging record all regions of their field of view in parallel, so $\eta_B\sqrt{A}$ compares them directly; the scanning penalty of a single-point probe discussed above applies to neither.

\begin{table}[ht]
\centering
\caption{Zero-field resonance through the bundle and in direct imaging of an ND layer \cite{Sengottuvel2025}, analyzed in the same way. $A$: area of the region ($3 \times 3$ binned pixels); $C$, $w$ and $\eta_B\sqrt{A}$ (Eq.~\eqref{eq:eta_slope}, exposure time) are medians over 360--400 regions in the quieter half of each map; values for the bundle at $40\times$ are means of two measurements. The last column scales the sensitivity to an LED current of \qty{0.9}{\ampere}, assuming that the photoluminescence is proportional to the current.}
\label{tab:freespace}
\footnotesize
\begin{tabular}{lcccccc}
\hline
Configuration & LED & $C$ & $w$ & $A$ & $\eta_B\sqrt{A}$ & at \qty{0.9}{\ampere} \\
 & (\unit{\ampere}) & (\%) & (\unit{\milli\tesla}) & (\unit{\micro\meter\squared}) & \multicolumn{2}{c}{(\unit{\milli\tesla\micro\meter\per\sqrt{\hertz}})} \\
\hline
Bundle, $20\times$ dry & 0.9 & 0.35 & 2.1 & 14 & 1.14 & 1.14 \\
Bundle, $40\times$ dry (NA 0.95) & 0.3 & 0.44 & 2.9 & 19 & 0.75 & 0.43 \\
Bundle, $40\times$ oil (NA 1.4) & 0.3 & 0.60 & 2.9 & 20 & 0.53 & 0.31 \\
Direct, $40\times$ dry (NA 0.65), 3 sweeps & 0.9 & 0.76 & 2.5 & 3.5 & 0.20 & 0.20 \\
Direct, $40\times$ dry (NA 0.65), 10 sweeps & 0.9 & 0.67 & 2.4 & 3.5 & 0.23 & 0.23 \\
\hline
\end{tabular}
\end{table}

\begin{table}[ht]
\centering
\caption{Sensitivity away from the trace, from the slope of the resonance and the measured noise (Eq.~\eqref{eq:eta_slope}), for spectra averaged over square regions of $n \times n$ binned pixels (\qty{1.24}{\micro\meter} each). $\delta S$: noise of the normalized signal per field step, with the shot noise for comparison. $\eta_B$ and $\eta_B\sqrt{A}$ are given for the total measurement time (\qty{31.9}{\second} per field step) and, in parentheses, for the exposure time alone (\qty{2.24}{\second} per field step). $\sigma_B$: uncertainty of the resonance position from fits of the whole sweep, as in the maps of Fig.~\ref{fig:endoscope}.}
\label{tab:sensitivity}
\small
\begin{tabular}{lcccc}
\hline
Region ($n \times n$) & $1 \times 1$ & $3 \times 3$ & $7 \times 7$ & $14 \times 14$ \\
Area $A$ (\unit{\micro\meter\squared}) & 1.5 & 14 & 76 & 302 \\
\hline
Contrast $C$ (\%) & 0.37 & 0.35 & 0.37 & 0.38 \\
Width $w$ (\unit{\milli\tesla}) & 2.2 & 2.2 & 2.2 & 2.2 \\
$I_0$ (\unit{\per\second}) & $1.0 \times 10^{6}$ & $9.0 \times 10^{6}$ & $4.9 \times 10^{7}$ & $2.0 \times 10^{8}$ \\
$\delta S$ ($10^{-4}$) & 13 & 4.7 & 2.2 & 1.0 \\
$\delta S$, shot noise ($10^{-4}$) & 13 & 4.4 & 1.9 & 0.95 \\
$\eta_B$ (\unit{\milli\tesla\per\sqrt{\hertz}}) & 3.0 (0.78) & 1.1 (0.30) & 0.53 (0.14) & 0.23 (0.060) \\
$\eta_B\sqrt{A}$ (\unit{\milli\tesla\micro\meter\per\sqrt{\hertz}}) & 3.7 (0.97) & 4.3 (1.1) & 4.6 (1.2) & 3.9 (1.04) \\
\hline
$\sigma_B$ (\unit{\micro\tesla}), whole sweep & 114 & 45 & 23 & 15 \\
Reduced $\chi^2$, whole sweep & 1.1 & 1.3 & 1.9 & 3.2 \\
\hline
\end{tabular}
\end{table}

\subsection{Dry and immersion readout of the magnetic measurement}
\label{sec:S_airoil}

\textbf{Measurements.} Four field maps of the test structure were recorded through the endoscope bundle on the same day: two with the dry $40\times$ objective ($\mathrm{NA} = 0.95$, UPLXAPO40X; exposure \qty{10}{\milli\second}) and two with the oil-immersion $40\times$ objective ($\mathrm{NA} = 1.4$, UPLXAPO40XO; exposure \qty{5}{\milli\second}), each pair with opposite directions of the current through the trace. The LED current (\qty{0.3}{\ampere}), the field sweep (80 steps from $-4.0$ to \qty{3.9}{\milli\tesla}, 16 sweeps) and the processing ($4 \times 4$ binning, reference/signal ratio, Gaussian fits of spectra averaged over $3 \times 3$ binned pixels) were the same as for Fig.~\ref{fig:endoscope}. The immersion measurements preceded the dry ones; residual oil remaining on the output facet would not change the dry readout, because guided light with $q > 1$ is then totally reflected at the oil--air interface.

\textbf{Registration and core sampling.} Each map was registered to one of the dry measurements by maximizing the correlation of the high-pass-filtered photoluminescence images, which are dominated by the core lattice. The two dry maps coincide (correlation 0.92); both immersion maps are scaled by 1.026, rotated by $0.2^\circ$ and shifted by eight binned pixels. A shift by one lattice vector lowers both the lattice correlation (0.74 to 0.58--0.62) and the correlation of the field maps (0.875 to 0.78--0.79), so that the registration is unambiguous. The cores were located as local maxima of the dry photoluminescence image (604 cores, 572 with valid fits in all four maps). For every core we averaged the photoluminescence over a disk of radius \qty{3.6}{\micro\meter} (core diameter \qty{10}{\micro\meter}), took the 20th percentile of an annulus at 6.5--\qty{9.2}{\micro\meter} as the inter-core level, and the median of the fitted parameters over the disk. The offset of the camera, estimated from the corners of the frames, whose signal scales with the exposure time, is about 2 ADU and was neglected.

\textbf{Corrections.} Ratios measured per camera pixel were converted to ratios per unit area of the facet with the square of the registration scale, 1.052. Both objectives are infinity-corrected and have the same focal length, so with the camera in the focal plane of the tube lens they would give the same magnification, independent of the distance between objective and tube lens. Here the camera was not in this plane (Methods), so the magnification depends slightly on the objective, which we take to be the origin of this small difference. The power delivered by the LED, which was focused onto the back focal plane of both objectives with equal focal lengths, differs by the transmission of the objectives at the LED wavelength, $T_\mathrm{oil}/T_\mathrm{dry} = 0.973$ (89\% and 91.5\% at \qty{520}{\nano\meter}, from the manufacturer's curves); the transmission of the photoluminescence differs by 0.96--0.98 (650--\qty{800}{\nano\meter}). The excitation enters the cores ($n = 1.908$) with a Fresnel transmission of 0.899 from air and 0.986 from oil, a factor of 1.097 in favour of the immersion readout. The Fresnel transmission of the photoluminescence at the output facet is part of the model (Supplementary Table~\ref{tab:readout}). Together, the measured ratios were divided by $1.052 \times 1.035 = 1.089$; the uncertainty of the transmission factor ($\pm 1\%$) is included in the quoted errors.

\begin{figure*}[ht]
\centering
\includegraphics[width=\textwidth]{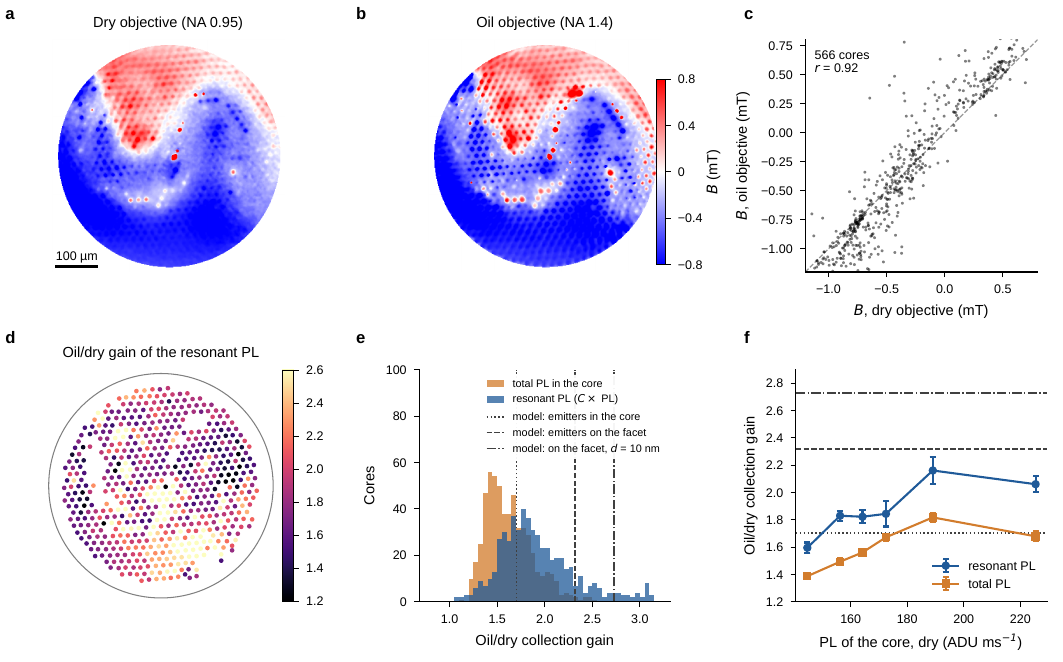}
\caption{\textbf{Magnetic-field imaging with dry and immersion readout.} \textbf{a},~\textbf{b},~Current-induced field (half difference of the maps recorded with opposite currents) through the dry ($\mathrm{NA} = 0.95$) and the oil-immersion ($\mathrm{NA} = 1.4$) $40\times$ objectives, in the coordinates of the dry measurement (not rotated to the orientation of Fig.~\ref{fig:endoscope}). Scale bar, \qty{100}{\micro\meter}. \textbf{c},~Field per core, immersion versus dry readout; dashed line, equality. \textbf{d},~Map of the immersion/dry gain of the resonant photoluminescence ($C \times \mathrm{PL}$) per core, corrected as in Table~\ref{tab:airoil}. \textbf{e},~Distributions of the gain of the resonant (blue) and of the total (orange) photoluminescence of the cores, with the model values for isotropic emission inside the core (dotted), emitters on the facet at $d = \qty{70}{\nano\meter}$ (dashed) and at \qty{10}{\nano\meter} (dash-dotted). \textbf{f},~Gain as a function of the photoluminescence of the core with dry readout (medians of six groups of equal size; error bars, standard errors of the medians).}
\label{fig:S_airoil}
\end{figure*}

\textbf{Results.} Table~\ref{tab:airoil} summarizes the ratios. All four combinations of a dry and an immersion measurement agree to within 2\% for the resonant signal; between the two measurements with the same objective, this signal is reproducible to 0.3\% (dry) and 1.7\% (oil), whereas the total photoluminescence of the dry measurements differs by 5\%. The gain of the resonant signal does not depend on the radial position in the bundle (1.97--2.09 before correction in four radial zones), but it increases with the brightness of the core, from 1.6 in the dimmest sixth of the cores to 2.1--2.2 in the brightest third (Supplementary Fig.~\ref{fig:S_airoil}f), which also show the highest contrast. The half differences of the maps recorded with opposite currents agree for the two objectives (Supplementary Fig.~\ref{fig:S_airoil}a--c; slope 1.02 for all pixels in the bundle, correlation 0.92 between cores), and the current-independent half sums are \qty{-0.04}{\milli\tesla} for both objectives, the residual laboratory field along the fiber axis. The differences between the two objectives (\qty{0.16}{\milli\tesla} robust standard deviation per core) exceed the statistical uncertainty of the fits (\qty{0.065}{\milli\tesla}); they are largest above the conductor, where the transverse field reduces the depth of the resonance, and indicate that the two objectives weight the emitters within a core differently.

\textbf{Sensitivity.} Normalized to the exposure time and to the area, the sensitivity (Eq.~\eqref{eq:eta_slope}) was 0.75 (dry) and \qty{0.53}{\milli\tesla\micro\meter\per\sqrt{\hertz}} (oil) for $3 \times 3$ binned pixels, compared with 1.14 for the $20\times$ measurement of Fig.~\ref{fig:endoscope} at three times the LED current (Table~\ref{tab:freespace}). The immersion objective is thus 1.4 times more sensitive than the dry one, which halves the measurement time needed for a given precision. Most of this gain comes from the contrast (a factor of 1.35) and only a smaller part from the photoelectron rate (a factor of 1.25 per unit facet area, or 1.12 in sensitivity): the additional light collected with immersion is ND emission rather than background, as expected for guided supercritical emission. For regions of the size of a core cell ($14 \times 14$ binned pixels, scaled to \qty{319}{\micro\meter\squared}), the sensitivity was 0.039 (dry) and \qty{0.035}{\milli\tesla\per\sqrt{\hertz}} (oil), compared with \qty{0.060}{\milli\tesla\per\sqrt{\hertz}} for the $20\times$ map (Table~\ref{tab:sensitivity}); the advantage of immersion shrinks here because, for large regions of the immersion maps, the noise exceeds the shot noise by about 25\%, whereas it stays within 10\% of it for the dry maps. Over the total measurement time, the three measurements give similar values (4--\qty{7}{\milli\tesla\micro\meter\per\sqrt{\hertz}}), because the camera was exposed for only 1.3\% (dry) and 0.7\% (oil) of the time. The $40\times$ maps appear noisier than Fig.~\ref{fig:endoscope}b only because their total exposure time (\qty{12.8}{\second} and \qty{25.6}{\second}) was 7 to 14 times shorter than that of the $20\times$ map (\qty{179}{\second}). Immersion readout of the whole bundle at full LED current and comparable integration time is therefore expected to give maps of better quality than Fig.~\ref{fig:endoscope}b.

\begin{table}[ht]
\centering
\caption{Ratios of the immersion ($\mathrm{NA} = 1.4$) to the dry ($\mathrm{NA} = 0.95$) readout for 572 cores. Raw: median over cores per camera pixel, mean $\pm$ standard deviation of the four combinations of measurements. Gain: per unit area of the facet, corrected for the transmission of the objectives and the Fresnel transmission of the excitation (factor 1.089 in total). Model values from Supplementary Table~\ref{tab:readout}.}
\label{tab:airoil}
\small
\begin{tabular}{lcc}
\hline
Quantity & Raw ratio & Collection gain \\
\hline
Resonant photoluminescence, $C \times \mathrm{PL}$ & $2.03 \pm 0.03$ & $1.86 \pm 0.03$ \\
Total photoluminescence of the cores & $1.72 \pm 0.05$ & $1.58 \pm 0.05$ \\
Photoluminescence between the cores & $1.33 \pm 0.04$ & $1.22 \pm 0.04$ \\
PL contrast $C$ & $1.21 \pm 0.04$ & -- \\
\hline
Model: emitters on the facet ($d = 70$ / \qty{10}{\nano\meter}) & & 2.32 / 2.73 \\
Model: isotropic emission inside the core & & 1.70 \\
\hline
\end{tabular}
\end{table}

\end{appendices}

\end{document}